\documentclass[twoside,11pt]{article}
\usepackage{a4wide,pictex,cite,latexsym,amsfonts,amssymb,exscale,epsfig,bbm}
\usepackage[centertags,sumlimits,intlimits,namelimits,reqno]{amsmath}
\usepackage{amsthm}

\theoremstyle{definition}
\newtheorem{theorem}{Theorem}[section]
\newtheorem{lemma}[theorem]{Lemma}
\newtheorem{corollary}[theorem]{Corollary}
\newtheorem{proposition}[theorem]{Proposition}
\newtheorem{definition}[theorem]{Definition}
\newtheorem{remark}[theorem]{Remark}

\def\href#1#2{#2}

\makeatletter
\newfont{\@aidxte}{cmsy10}
\newfont{\@aidxel}{cmsy10 scaled 1095}
\newfont{\@aidxtw}{cmsy10 scaled 1200}
\newlength\@aidxtexvi
\newlength\@aidxtexvii
\newlength\@aidxelxvi
\newlength\@aidxelxvii
\newlength\@aidxtwxvi
\newlength\@aidxtwxvii
\newcommand{\alignidx}[1]{%
  \@aidxtexvi=\fontdimen16\@aidxte
  \@aidxtexvii=\fontdimen17\@aidxte
  \@aidxelxvi=\fontdimen16\@aidxel
  \@aidxelxvii=\fontdimen17\@aidxel
  \@aidxtwxvi=\fontdimen16\@aidxtw
  \@aidxtwxvii=\fontdimen17\@aidxtw
    {\mbox{$%
    \fontdimen16\@aidxte=2.9pt
    \fontdimen17\@aidxte=2.9pt
    \fontdimen16\@aidxel=3.1pt
    \fontdimen17\@aidxel=3.1pt
    \fontdimen16\@aidxtw=3.3pt
    \fontdimen17\@aidxtw=3.3pt
    #1$}}%
    \fontdimen16\@aidxte=\@aidxtexvi
    \fontdimen17\@aidxte=\@aidxtexvii
    \fontdimen16\@aidxel=\@aidxelxvi
    \fontdimen17\@aidxel=\@aidxelxvii
    \fontdimen16\@aidxtw=\@aidxtwxvi
    \fontdimen17\@aidxtw=\@aidxtwxvii}

\@ifundefined{chapter}{%
  \renewcommand{\theequation}{\thesection.\arabic{equation}}%
  \@addtoreset{equation}{section}}{%
  \renewcommand{\theequation}{\thechapter.\arabic{equation}}%
  \@addtoreset{equation}{chapter}}%

\newcounter{mathletter}%
\newcommand{\bmathletter}{%
  \refstepcounter{equation}%
  \setcounter{mathletter}{\value{equation}}%
  \setcounter{equation}{0}%
  \@ifundefined{chapter}{%
    \renewcommand{\theequation}{%
      \mbox{\thesection.\arabic{mathletter}\alph{equation}}}}{%
    \renewcommand{\theequation}{%
      \mbox{\thechapter.\arabic{mathletter}\alph{equation}}}}}%
\newcommand{\emathletter}{\setcounter{equation}{\value{mathletter}}}%
\newenvironment{mathletters}{\bmathletter}{\emathletter}%
\makeatother

\newenvironment{myenumerate}{%
  \begin{enumerate}
  \setlength{\partopsep}{0pt}
  \setlength{\parskip}{0pt}}{\end{enumerate}}

\newcommand{\hpeprint}[1]{%
  \href{http://arXiv.org/abs/#1}{\texttt{#1}}}%
\newcommand{\hpspires}[1]{}%
\def\pacs#1{\noindent PACS: #1\par}%
\def\keywords#1{\noindent key words: #1\par}%
\def\acknowledgements{\section*{Acknowledgements}}%
\def\draft#1{}
\def\dopreprint{\hfill{\small\thepreprint}\\}%
\def\preprint#1{\def\thepreprint{#1}}%
\def\thepreprint#1{}%
\def\sym#1{{\mathcal #1}}
\def\emph#1{{\sl #1\/}}
\newcommand{\ontop}[2]{\genfrac{}{}{0pt}{2}{\scriptstyle #1}{\scriptstyle #2}}
\let\phi=\varphi
\let\theta=\vartheta
\let\epsilon=\varepsilon
\let\hat=\widehat
\let\tilde=\widetilde
\def\address#1{\date{{\sl #1}\\\ \\\theversion}\gdef\date##1{}}%
\def\version#1{\gdef\theversion{#1}}%
\def\mycaption#1#2{%
  \begin{quote}
  \caption{\label{#1}#2}
  \end{quote}}
\def\nn{\notag}
\def\eqref#1{(\ref{#1})}%

\def\tr{\mathop{\rm tr}\nolimits}

\def\dim{\mathop{\rm dim}\nolimits}

\def\id{\mathop{\rm id}\nolimits}

\def\Hom{\mathop{\rm Hom}\nolimits}
\def\Aut{\mathop{\rm Aut}\nolimits}

\def\qdim{\mathop{\rm qdim}\nolimits}
\def\qtr{\mathop{\rm qtr}\nolimits}
\def\CC{\sym{C}}
\def\RCat{\mathop{\rm Rep}\nolimits}
\def\Rep{\tilde{\sym{R}}}%
\def\Irrep{\sym{R}}%
\def\del{\partial}
\def\S{\sym{S}}

\def\ev{\mathop{\rm ev}\nolimits}
\def\coev{\mathop{\rm coev}\nolimits}
\def\sgn{\mathop{\rm sgn}\nolimits}
\def\1{\mathbf{1}}

\def\Calg{C_{\rm alg}}%

\def\SO{{SO}}

\def\SU{{SU}}

\def\ie{{\sl i.e.\/}}

\def\etc{{\sl etc.\/}}
\def\cf{{\sl cf.\/}}

\def\C{{\mathbbm C}}
\def\N{{\mathbbm N}}
\def\R{{\mathbbm R}}
\def\Z{{\mathbbm Z}}

\def\g{{\mathfrak{g}}}

\version{6 September 2001}%
\preprint{DAMTP-2001-37}

\begin{document}
%

\title{\dopreprint Four-dimensional Lattice Gauge Theory with ribbon categories\\ 
       and the Crane--Yetter state sum}
\author{Hendryk Pfeiffer\thanks{e-mail: H.Pfeiffer@damtp.cam.ac.uk}}
\address{Department of Applied Mathematics and Theoretical Physics,\\
         Centre for Mathematical Sciences,\\
         Wilberforce Road,\\ 
         Cambridge CB3 0WA, UK}
\date{\version}
\maketitle

%
\begin{abstract}
%

  Lattice Gauge Theory in $4$-dimensional Euclidean space-time is
  generalized to ribbon categories which replace the category of
  representations of the gauge group. This provides a framework in
  which the gauge group becomes a quantum group while space-time is
  still given by the `classical' lattice. At the technical level, this
  construction generalizes the Spin Foam Model dual to Lattice Gauge
  Theory and defines the partition function for a given triangulation
  of a closed and oriented piecewise-linear $4$-manifold. This
  definition encompasses both the standard formulation of $d=4$ pure
  Yang--Mills theory on a lattice and the Crane--Yetter invariant of
  $4$-manifolds. The construction also implies that a certain class of
  Spin Foam Models formulated using ribbon categories are well-defined
  even if they do not correspond to a Topological Quantum Field
  Theory.
\end{abstract}



%
\section{Introduction}
%

The formulation of gauge theory on a lattice~\cite{Wi74} combines
manifest gauge symmetry with the path integral approach although
space-time cannot be retained as a smooth manifold and is replaced
instead by a discrete structure. In the present paper Lattice Gauge
Theory (LGT) always refers to pure gauge theory in Euclidean
space-time.

LGT offers a number of generalizations that do not have a na{\"\i}ve
continuum analogy such as gauge theory with finite gauge
groups. Furthermore in three dimensions it is possible to define LGT
for quantum groups~\cite{Bo93,Oe01}. Combining the various actions and
Boltzmann weights with suitable `gauge groups' (finite groups, Lie
groups or quantum groups), this model has several special cases that
belong to different branches of physics and mathematics. It is at the
centre of the relation between LGT with Yang--Mills~\cite{Wi74} or
with Chern--Simons action~\cite{DiWi90,Wa92}, the Turaev--Viro
invariant of $3$-manifolds~\cite{TuVi92,BaWe96}, a purely algebraic
construction of Topological Quantum Field Theory~\cite{DiWi90,TuVi92}
and $3$-dimensional Euclidean quantum gravity without or with
cosmological constant~\cite{Ba99}.

At least some of the above constructions are known to have analogies
in four dimensions. Even though the question of which is a suitable
unified model remains unsolved in full generality, some of the
relations known from three dimensions persist also in four
dimensions. In the present paper we concentrate one the standard
formulation of LGT and on the Crane--Yetter state
sum~\cite{CrYe93,CrKa97}. We present a definition which encompasses
both and generalizes four-dimensional LGT to quantum
groups. Technically this is realized for ribbon categories which arise
as the categories of representations of certain quantum groups and
which replace the category of representations of the gauge group of
LGT.

The main result of the present paper is the existence of such a
generalized LGT in four-dimensional Euclidean space-time using ribbon
categories. This model contains the Crane--Yetter state sum as a
special case for a particular Boltzmann weight and agrees on the other
hand with the Spin Foam Model which is strong-weak dual to LGT if the
ribbon category is the category of finite-dimensional representations
of a compact Lie group. Beyond this, it provides a definition of Spin
Foam Models in $d=4$ using ribbon categories which includes in
particular the proof that this construction is well-defined even in
cases in which the model does not correspond to a Topological Quantum
Field Theory.

At the technical level, the construction of these Spin Foam Models
using ribbon categories can be motivated from the following
observations.  From the study of non-perturbative quantum gravity it
has emerged that LGT admits a reformulation as a Spin Foam Model ---
see, for example~\cite{ReRo97,Ba98,Ba99}. Many models of interest in
quantum gravity are either Topological Quantum Field Theories and use
`delta-functions' as Boltzmann weights (for example~\cite{Oo92}), or
they are topological up to constraints which do not change the
weights, but restrict the set of admissible representations. This is
the case for some versions of the Barrett--Crane model~\cite{BaCr98}.

LGT, however, admits more general Boltzmann weights,
\begin{equation}
\label{eq_boltzmann}
  w\colon G\to\R,\qquad g\mapsto \exp(-s(g)).
\end{equation}
Here the compact Lie group $G$ is the gauge group, the (local) action
$s\colon G\to\R$ is a real, bounded and $L^2$-integrable class
function, and the Boltzmann weight $w(g)$ is evaluated for each
plaquette of the lattice. This model encompasses lattice Yang--Mills
theory, for example using Wilson's action, but it is not restricted to
this case. For general background on LGT the reader is referred to
standard textbooks, for example~\cite{MoMu94,Ro92}.

The Spin Foam Model corresponding to the standard formulation of LGT
on a hypercubic lattice was constructed in detail in~\cite{OePf01}
where it was found that it generalizes the strong-weak dual of LGT
which had been known only in the Abelian case~\cite{Pe78,Sa80} and for
$\SU(2)$ in $d=3$~\cite{AnCh93} before. The Boltzmann
weight~\eqref{eq_boltzmann} enters the Spin Foam Model via the
coefficients $\hat w_\rho$ of its character expansion,
\begin{equation}
\label{eq_charexp}
  w(g)=\sum_{\rho\in\Irrep}\hat w_\rho\,\chi^{(\rho)}(g),\qquad
  \hat w_\rho = \dim V_\rho\,\int_G\overline{\chi^{(\rho)}(g)}w(g)\,dg.
\end{equation}
Here $\chi^{(\rho)}\colon G\to\C$ denotes the character of the
finite-dimensional irreducible representation $V_\rho$, the sum is
over equivalence classes of finite-dimensional irreducible
representations of $G$, and $\int_G$ is the normalized Haar
measure on $G$.

The way the coefficients $\hat w_\rho$ appear in the Spin Foam Model
dual to LGT~\cite{OePf01} compared with the Ooguri state
sum~\cite{Oo92} indicates that there exists a unified construction
encompassing both. In addition, the fact that the Crane--Yetter state
sum~\cite{CrYe93,CrKa97} can be understood as a generalization of the
Ooguri model to quantum groups, suggests the construction given in the
present paper.

The strategy for the definition of $d=4$ LGT using ribbon categories
is as follows. The construction is based on a triangulation of a
closed and oriented piecewise-linear four-manifold $M$ which is
specified by an abstract combinatorial complex. In the special case of
a Lie group symmetry, the definition shall coincide with the Spin Foam
Model dual to LGT if that LGT is formulated on the $2$-complex dual to
the triangulation (note that we formulate the Spin Foam Model on the
triangulation itself following~\cite{CrKa97}). In the Lie group case,
both pictures are available: the Spin Foam Model on the triangulation
and LGT on the dual $2$-complex. They are dual to each other in the
sense of~\cite{OePf01}. Physically this means strong-weak duality
between LGT and the Spin Foam Model while on the mathematical side the
two models are related by a Tannaka--Krein like reconstruction theorem
relating LGT (formulated in terms of the gauge group $G$) with the
spin foam model (formulated in terms of the category of
representations $\RCat G$). For details on quantum groups, ribbon
categories and the reconstruction theorems, the reader is referred to
standard textbooks such as \cite{ChPr94,Ma95a}.

The generalization takes place in the spin foam picture where the
category $\RCat G$ is replaced by a suitable ribbon category
$\CC$. Loosely speaking, using the reconstruction theorems, this
provides a definition of LGT in which the gauge group is replaced by a
quantum group. Technically, the notion of gauge group is lost, but one
can think of replacing the algebra of representation functions
$\Calg(G)$ by a non-commutative algebra (a suitable ribbon Hopf
algebra) while space-time is still given by the `classical' lattice.

The generalization from $\RCat G$ to a generic ribbon category $\CC$
involves choices of the ordering of tensor factors and choices of the
braiding whenever tensor factors are exchanged. These choices are not
at all obvious from the Lie group case which involves only the
symmetric category $\RCat G$.

The method to achieve a consistent definition in the Spin Foam picture
is to choose a linear order of vertices for the combinatorial complex
and to define the partition function in a way that employs special
choices and that refers explicitly to that order. It is then possible
to show in a second step that the partition function is actually
independent of the order (combinatorially invariant) and is thus
well-defined for a given triangulation. This approach can be seen as a
generalization to four dimensions of the strategy which Barrett and
Westbury~\cite{BaWe96} employ in their approach to the Turaev--Viro
invariant~\cite{TuVi92}.

Another point of view on the definition given in the present paper is
related to the construction of the Crane--Yetter state sum
in~\cite{CrKa97}. The authors of~\cite{CrKa97} first show that the
state sum is independent of the triangulation which in our terminology
relies on the choice of a particular Boltzmann weight. Triangulation
independence then implies combinatorial invariance and thus
establishes that the state sum is well-defined. As an alternative
proof, it is conceivable to show combinatorial invariance in the first
step. This holds for any choice of Boltzmann weights. One could then
prove in a second step that the choice of special Boltzmann weights
implies triangulation independence by standard arguments as
in~\cite{Oo92,CrYe93,CrKa97}. The construction presented in the
present paper can be viewed as the first of these two steps.

Finally, we would like to mention D.~V.~Boulatov's approach to LGT for
quantum groups in $3$ dimensions~\cite{Bo93}. His construction makes
use of the general result of Reshetikhin and Turaev~\cite{ReTu90}
establishing a functor from the category of ribbon graphs in $\R^3$ to
the ribbon category $\CC$. The strategy in~\cite{Bo93} is to construct
a suitable ribbon graph in the triangulated manifold which then yields
a well-defined partition function as the quantum trace of a ribbon
morphism.

A related definition of $d=3$ LGT for quantum groups was developed by
R.~Oeckl~\cite{Oe01} in which the duality between LGT and its dual
Spin Foam Model is understood entirely in terms of manipulations of
ribbon graphs. The approach of~\cite{Oe01} also develops the
correspondence of ribbon categories with suitable quantum groups,
namely coribbon Hopf algebras, in a way that transparently generalizes
the duality transformation of the Lie group case.

However, since the Reshetikhin--Turaev functor is available only for
ribbon graphs in $\R^3$, these approaches do not have a direct
generalization to higher dimension. In the present paper, we use the
functor mainly to justify diagrammatic calculations.

The present paper is organized as follows. In
Section~\ref{sect_prelim}, we review some mathematical background on
the Peter--Weyl theory for compact Lie groups and on ribbon
categories, and we introduce our notation for combinatorial and
simplicial complexes. The duality transformation for LGT with Lie
gauge groups which was derived in~\cite{OePf01} on a cubic lattice is
reviewed in Section~\ref{sect_duality} and formulated there on a
$2$-complex.  In Section~\ref{sect_ribbonlgt}, we define the Spin Foam
Model generalizing the dual of LGT to suitable ribbon categories. This
section contains the definition of the partition function, the proof
that it is well-defined and comments on the construction of
observables and on the role played by the gauge transformations in the
Spin Foam Model. In Section~\ref{sect_special}, we indicate how these
definitions specialize to the standard formulation of LGT with a
compact Lie group (or a finite group) as the gauge group and to the
Crane--Yetter invariant. We also comment on possible generalizations
and relations with other spin foam models. Section~\ref{sect_outlook}
contains a conclusion and comments on open questions.

%
\section{Preliminaries}
%
\label{sect_prelim}

\subsection{Peter--Weyl Theory}

In this section, we briefly summarize definitions and basic statements
related to the algebra of representation functions $\Calg(G)$ of $G$
where $G$ is a compact Lie group (or a finite group). These results
are needed in Section~\ref{sect_duality} in order to present the
duality transformation relating LGT and the Spin Foam Model. For more
details, the reader is referred to the introduction of~\cite{OePf01}
or to textbooks such as~\cite{BrDi85,CaSe95}.

\subsubsection{Representation functions}

Finite-dimensional complex vector spaces on which $G$ is represented
are denoted by $V_\rho$ and by $\rho\colon G\to\Aut V_\rho$ the
corresponding group homomorphism. Let $\Rep$ denote a set containing
one unitary representative of each class of finite-dimensional
representations and $\Irrep$ the subset of irreducible
representations. For a representation $\rho\in\Rep$, the dual
representation is denoted by $\rho^\ast$, and the dual vector space of
$V_\rho$ by $V_\rho^\ast$. The dual representation is given by
$\rho^\ast\colon G\mapsto \Aut V_\rho^\ast$, where $\rho^\ast(g)\colon
V_\rho^\ast\to V_\rho^\ast$, $\eta\mapsto\eta\circ\rho(g^{-1})$, \ie\
$(\rho^\ast(g)\eta)(v)=\eta(\rho(g^{-1})v)$ for all $v\in V_\rho$.
There exists a one-dimensional `trivial' representation of $G$ which
is denoted by $V_{[1]}\cong\C$.

For the unitary representations $V_\rho$, $\rho\in\Rep$, there exist
standard sesquilinear scalar products $\left<\cdot;\cdot\right>$ and
orthonormal bases $(v_j)$ in such a way that the basis $(v_j)$ of
$V_\rho$ is dual to the basis $(\eta^j)$ of $V_\rho^\ast$, \ie\
$\eta^j(v_k)=\delta^j_k$. Duality is here given by the scalar
product, \ie\ $\left<v_j;v_k\right>=\eta^j(v_k)$ and
$\bigl<\eta^j;\eta^k\bigr>=\eta^k(v_j)$, $1\leq j,k\leq\dim V_\rho$.

The complex-valued functions
\begin{equation}
  t_{\eta,v}^{(\rho)}\colon G\to\C,\quad g\mapsto\eta(\rho(g)v),
\end{equation}
where $\rho\in\Rep$, $v\in V_\rho$ and $\eta\in V_\rho^\ast$, are
called \emph{representation functions} of $G$. They form a commutative
and associative unital algebra over $\C$,
\begin{equation}
  \Calg(G) := \{\,t_{\eta,v}^{(\rho)}\colon\quad
    \rho\in\Rep, v\in V_\rho, \eta\in V_\rho^\ast\,\},
\end{equation}
whose operations are given by
\begin{mathletters}
\begin{eqnarray}
  (t_{\eta,v}^{(\rho)} + t_{\theta,w}^{(\sigma)})(g)
    &:=& t_{\eta+\theta,v+w}^{(\rho\oplus\sigma)}(g),\\
  (t_{\eta,v}^{(\rho)}\cdot t_{\theta,w}^{(\sigma)})(g)
    &:=& t_{\eta\otimes\theta,v\otimes w}^{(\rho\otimes\sigma)}(g),
\end{eqnarray}%
\end{mathletters}%
for $\rho,\sigma\in\Rep$ and $v\in V_\rho$, $w\in V_\sigma$,
$\eta\in V_\rho^\ast$, $\theta\in V_\sigma^\ast$ and $g\in G$. The
zero element of $\Calg(G)$ is $t_{0,0}^{[1]}(g)=0$ and its
unit element $t_{\eta,v}^{[1]}(g)=1$ where the normalization
is such that $\eta(v)=1$.

The algebra $\Calg(G)$ is furthermore equipped with a Hopf algebra
structure employing the coproduct
$\Delta\colon\Calg(G)\to\Calg(G)\otimes\Calg(G)\cong\Calg(G\times G)$,
the co-unit $\epsilon\colon\Calg(G)\to\C$ and the antipode
$S\colon\Calg(G)\to\Calg(G)$ which are defined by
\begin{mathletters}
\begin{eqnarray}
\label{eq_matrixcopro}
  \Delta t_{\eta,v}^{(\rho)} (g,h) 
    &:=& t_{\eta,v}^{(\rho)}(g\cdot h),\\
  \epsilon t_{\eta,v}^{(\rho)}
    &:=& t_{\eta,v}^{(\rho)}(1),\\
\label{eq_matrixanti}
  S t_{\eta,v}^{(\rho)} (g) 
    &:=& t_{\eta,v}^{(\rho)}(g^{-1}),
\end{eqnarray}%
\end{mathletters}%
for $\rho\in\Rep$ and $v\in V_\rho$, $\eta\in V_\rho^\ast$ and
$g,h\in G$. For unitary representations, the antipode relates a
representation with its dual which is just the conjugate
representation,
\begin{equation}
\label{eq_antipode}
  S t_{mn}^{(\rho)}(g) 
    = t_{nm}^{(\rho^\ast)}(g)
    = \overline{t_{nm}^{(\rho)}(g)}.
\end{equation}
The bar denotes complex conjugation.

\subsubsection{Peter--Weyl decomposition and theorem}

The structure of the algebra $\Calg(G)$ can be understood if
$\Calg(G)$ is considered as a representation of $G\times G$ by
combined left and right translation of the function argument.

\begin{theorem}[Peter--Weyl decomposition]
Let $G$ be a compact Lie group (or a finite group).
\begin{myenumerate}
\item 
  There is an isomorphism
\begin{equation}
\label{eq_structure_calg}
  \Calg(G)\cong_{G\times G} 
    \bigoplus_{\rho\in\Irrep}(V_\rho^\ast\otimes V_\rho),
\end{equation}
  of representations of $G\times G$. Here the direct sum is over one
  unitary representative of each equivalence class of
  finite-dimensional irreducible representations of $G$. The direct
  summands $V_\rho^\ast\otimes V_\rho$ are irreducible as representations
  of $G\times G$. 
\item
  The direct sum in~\eqref{eq_structure_calg} is orthogonal with
  respect to the $L^2$-scalar product on $\Calg(G)$ which is formed
  using the Haar measure on $G$ on the left hand side, and using the
  standard scalar products on the right hand side, namely
\begin{equation}
  {\bigl<t_{\eta,v}^{(\rho)};t_{\theta,w}^{(\sigma)}\bigr>}_{L^2}
    = \int_G\overline{t_{\eta,v}^{(\rho)}(g)}\cdot t_{\theta,w}^{(\sigma)}(g)\,dg
    = \frac{1}{\dim V_{\rho}}\delta_{\rho\sigma}
        \overline{\left<\eta;\theta\right>}\left<v;w\right>,
\end{equation}
  where $\rho,\sigma\in\Irrep$ are irreducible. The Haar measure is
  denoted by $\int_G$ and normalized such that $\int_G\,dg=1$.
\end{myenumerate}
\end{theorem}

Each representation function $f\in\Calg(G)$ can thus be decomposed
according to~\eqref{eq_structure_calg} such that its $L^2$-norm is
given by
\begin{equation}
  {||f||}^2_{L^2} = \sum_{\rho\in\Irrep}\frac{1}{\dim V_\rho}{||f_\rho||}^2,
\end{equation}
where $f_\rho\in V_\rho^\ast\otimes V_\rho\cong\Hom(V_\rho,V_\rho)$,
$\rho\in\Irrep$, and all except finitely many $f_\rho$ are zero.

\begin{theorem}[Peter--Weyl theorem]
Let $G$ be a compact Lie group. Then $\Calg(G)$ forms a dense subset
of $L^2(G)$.
\end{theorem}

The characters $\chi^{(\rho)}\colon G\to\C$ associated with the
finite-dimensional unitary representations $\rho\in\Rep$ of $G$ are
given by the traces,
\begin{equation}
  \chi^{(\rho)} := \sum_{j=1}^{\dim V_\rho} t_{jj}^{(\rho)}.
\end{equation}
Each class function $f\in\Calg(g)$ can be character-decomposed
\begin{equation}
  f(g) = \sum_{\rho\in\Irrep}\chi^{(\rho)}(g)\,\hat f_\rho,
    \qquad\mbox{where}\qquad 
  \hat f_\rho = \dim V_\rho\,\int_G\overline{\chi^{(\rho)}(g)}f(g)\,dg,
\end{equation}
such that the completion of $\Calg(G)$ to $L^2(G)$ is compatible with
this decomposition.

\subsubsection{The Haar measure}

The Haar measure on $G$ can be understood in terms of the Peter--Weyl
decomposition~\eqref{eq_structure_calg} as follows.

\begin{proposition}
Let $G$ be a compact Lie group (or a finite group) and $\rho\in\Rep$ be
a finite-dimensional unitary representation of $G$ with the orthogonal
decomposition
\begin{equation}
  V_\rho\cong\bigoplus_{j=1}^k V_{\tau_j},\qquad \tau_j\in\Irrep, k\in\N,
\end{equation}
into irreducible components $\tau_j$. Let $P^{(j)}\colon V_\rho\to
V_{\tau_j}\subseteq V_\rho$ be the $G$-invariant orthogonal projectors
associated with the above decomposition. Assume that precisely the
first $\ell$ components $\tau_1,\ldots,\tau_\ell$, $0\leq\ell\leq k$,
are equivalent to the trivial representation. Then the Haar measure
of a representation function $t_{mn}^{(\rho)}$, $1\leq m,n\leq\dim
V_\rho$, is given by
\begin{equation}
\label{eq_haaralg}
  \int_G t_{mn}^{(\rho)}(g)\,dg 
    = \sum_{j=1}^\ell P^{(j)m}P^{(j)}_n,\qquad
  P^{(j)m}=\eta^m(w^{(j)}),\quad
  P^{(j)}_n=\theta^{(j)}(v_n).
\end{equation}
Here $(v_n)$ and $(\eta^m)$ are dual bases of $V_\rho$ and
$V_\rho^\ast$, the $w^{(j)}$ are normalized vectors in
$V_{\tau_j}\subseteq V_\rho$, and $\theta^{(j)}$ denotes the 
linear form dual to $w^{(j)}$.
\end{proposition}

\subsection{Ribbon Categories}
\label{sect_ribbon}

Ribbon categories are used in the present paper as a generalization of
the category of representations of the gauge group. A ribbon category
is a braided monoidal category with some additional structure. In this
section, we summarize the basic definitions with emphasis on a
convenient diagrammatic notation. We refer the reader to the
literature for more details, for example, to the
textbooks~\cite{ChPr94,Ma95a}. Our presentation is similar to that
of~\cite{Oe01}; we essentially follow~\cite{Ma95a}, but use the
diagrams of~\cite{ChPr94}. Also relevant in the context of the present
paper are the results of Reshetikhin and Turaev~\cite{ReTu90,Tu94}.

\subsubsection{Basic definitions}

\begin{definition}
\label{def_monoidal}
A \emph{strict monoidal category} is a category $\CC$ together with a
covariant functor $\otimes\colon\CC\times\CC\to\CC$ and a \emph{unit
object} $\1$ such that
\begin{mathletters}
\begin{gather}
  U\otimes(V\otimes W)=(U\otimes V)\otimes W,\\
\label{eq_unitobject}
  V\otimes\1=V=\1\otimes V,
\end{gather}%
\end{mathletters}%
for all objects $U,V,W$.
\end{definition}

\begin{definition}
A \emph{strict braided monoidal category} is a strict monoidal
category with natural isomorphisms (the \emph{braiding}),
\begin{equation}
\label{eq_braiding}
  \psi_{V,W}\colon V\otimes W\to W\otimes V,
\end{equation}
such that
\begin{mathletters}
\begin{gather}
  \psi_{U\otimes V,W}=(\psi_{U,W}\otimes\id_V)\circ(\id_U\otimes\psi_{V,W}),\\
  \psi_{U,V\otimes W}=(\id_V\otimes\psi_{U,W})\circ(\psi_{U,V}\otimes\id_W),\\
\label{eq_trivialbraid}
  \psi_{V,\1}=\id_V=\psi_{\1,V},
\end{gather}%
\end{mathletters}%
for all objects $U,V,W$. The category is called \emph{symmetric} if in
addition
\begin{equation}
  \psi_{W,V}\circ\psi_{V,W}=\id_{V\otimes W}.
\end{equation}
\end{definition}

\begin{definition}
\label{def_rigid}
A strict monoidal category $\CC$ is called \emph{rigid} if for each
object $V$ there exists an object $V^\ast$ (the \emph{left dual}) and
if there are natural isomorphisms
\begin{mathletters}
\label{eq_evcoev}
\begin{alignat}{2}
  \ev_V  &\colon V^\ast\otimes V\to\1,&
         &\qquad\mbox{(evaluation)}\\
  \coev_V&\colon\1\to V\otimes V^\ast,&
         &\qquad\mbox{(co-evaluation)}
\end{alignat}%
\end{mathletters}%
which satisfy for all objects $V$,
\begin{mathletters}
\begin{eqnarray}
\label{eq_straighta}
  \id_V        &=&(\id_V\otimes\ev_V)\circ(\coev_V\otimes\id_V),\\
\label{eq_straightb}
  \id_{V^\ast} &=&(\ev_V\otimes\id_{V^\ast})\circ(\id_{V^\ast}\otimes\coev_V).
\end{eqnarray}%
\end{mathletters}%
For a given morphism $f\colon V\to W$, the dual morphism $f^\ast\colon
W^\ast\to V^\ast$ is defined by
\begin{equation}
\label{eq_dualmorphism}
  f^\ast := (\ev_W\otimes\id_{V^\ast})
           \circ(\id_{W^\ast}\otimes f\otimes\id_{V^\ast})
           \circ(\id_{W^\ast}\otimes\coev_V).
\end{equation}
Left duality thus defines a contravariant functor
$\ast\colon\CC\to\CC$.
\end{definition}

\begin{definition}
\label{def_ribbon}
A \emph{strict ribbon category} $\CC$ is a strict rigid braided
monoidal category with natural isomorphisms (the \emph{twist}),
\begin{equation}
\label{eq_twist}
  \nu_V\colon V\to V,
\end{equation}
such that for all objects $V,W$,
\begin{mathletters}
\begin{gather}
\label{eq_ribbontensor}
\nu_{V\otimes W} = (\nu_V\otimes\nu_W)\circ\psi_{W,V}\circ\psi_{V,W},\\
\label{eq_ribbondual}
{(\nu_V)}^\ast   = \nu_{V^\ast},\\
\label{eq_trivialtwist}
\nu_\1           = \id_\1.
\end{gather}%
\end{mathletters}%
\end{definition}

It is now possible to construct right duals ${}^\ast V$ from the
braiding, the twist and the left duals. The right dual objects agree
in this case with the left duals, ${}^\ast V=V^\ast$, and right
evaluation and right co-evaluation are given by,
\begin{mathletters}
\label{eq_rightdual}
\begin{alignat}{2}
  \tilde\ev_V&\colon V\otimes V^\ast\to\1,&\qquad
    \tilde\ev_V&:=\ev_V\circ\psi_{V,V^\ast}\circ(\nu_V\otimes\id_{V^\ast}),\\
  \tilde\coev_V&\colon\1\to V^\ast\otimes V,&\qquad
    \tilde\coev_V&:=(\nu_{V^\ast}\otimes\id_V)\circ\psi_{V,V^\ast}\circ\coev_V.
\end{alignat}%
\end{mathletters}%
Finally, right and left duals can be employed in order to define
the analogues of trace and dimension.

\begin{definition}
Let $\CC$ be a strict ribbon category, $V$ an object of $\CC$ and
$f\colon V\to V$.
\begin{myenumerate}
\item
  The \emph{quantum trace} of $f$ is defined by
\begin{equation}
\label{eq_qtrace}
  \qtr(f) := \tilde\ev_V\circ(f\otimes\id_{V^\ast})\circ\coev_V.
\end{equation}
\item
  The \emph{quantum dimension} of $V$ is defined by
\begin{equation}
\label{eq_qdim}
  \qdim V:= \qtr (\id_V) = \tilde\ev_V\circ\coev_V.
\end{equation}
\end{myenumerate}
\end{definition}

Note that the quantum trace satisfies $\qtr(g\circ f)=\qtr(f\circ g)$
for $f\colon V\to W$ and $g\colon W\to V$. Furthermore, for $h\colon
V\to V$ and $k\colon W\to W$, $\qtr(h\otimes k)=\qtr(h)\circ\qtr(k)$
and $\qdim (V\otimes W)=\qdim V\circ\qdim W$, where the compositions
are in $\Hom(\1,\1)$.

All monoidal categories defined above, starting from
Definition~\ref{def_monoidal}, are strict. If a non-strict category is
given, there exists an equivalent strict version~\cite{ReTu90} which
can be used instead. As a consequence of the coherence conditions on
associativity and unit constraints in the definition of a (non-strict)
monoidal category, it would also be possible to make a choice of
parentheses in all definitions and to insert the constraints in a
consistent way in all equations. The same would apply to the
calculations and results presented in the following sections of this
paper.

Furthermore, all categories of interest in this paper are $\C$-linear
(for details see, for example, \cite{ChPr94,Ma73}). This means that
there is the notion of a (finite) direct sum of objects, that
furthermore for given objects $V,W$ the sets $\Hom(V,W)$ form
$\C$-vector spaces and that composition of morphisms is
$\C$-bilinear. Additionally, there are notions of monomorphism and
epimorphism which have the usual properties known from linear
algebra. The reader might think of the case where all objects are
$\C$-vector spaces. Finally, the additional structures such as tensor
product, braiding, duality and twist are required to be compatible
with the $\C$-linear structure, in particular $\Hom(\1,\1)\cong\C$
such that composition corresponds to multiplication.

As a consequence, $\Hom(U\otimes V,W)\cong\Hom(V,W\otimes U^\ast)$ are
isomorphic as $\C$-vector spaces. We also need the dual space of
$\Hom(V,W)$. One can make use of a non-degenerate $\C$-bilinear
pairing
\begin{equation}
\label{eq_pairing}
  \Hom(V^\ast,W^\ast)\otimes_\C\Hom(V,W)\to\C,
  f\otimes g\mapsto \ev_W\circ(g\otimes f)\circ\tilde\coev_V,
\end{equation}
in order to define the dual space $\Hom(V,W)^\ast$ up to
isomorphism. Here we use $\Hom(V^\ast,W^\ast)$ rather than $\Hom(W,V)$
because some diagrams in the following sections are then related by a
mirror symmetry.

All conditions that are required for the construction of the Spin Foam
Model are summarized in the following definition.

\begin{definition}
\label{def_admissible}
An \emph{admissible} ribbon category is a $\C$-linear strict ribbon
category which satisfies the following conditions,
\begin{myenumerate}
\item
  For all objects $V$, $W$ of $\CC$ the space $\Hom(V,W)$ is
  finite-dimensional as a $\C$-vector space.
\item
  The pairing~\eqref{eq_pairing} is non-degenerate for all objects
  $V$, $W$ of $\CC$.
\end{myenumerate}
A \emph{set of colours} $\CC_0$ is a countable set of objects of
$\CC$ such that
\begin{myenumerate}
\item
  No two elements of $\CC_0$ are isomorphic,
\item
  For each object $V\in\CC_0$, the set $\CC_0$ also contains an object
  which is isomorphic to $V^\ast$.
\end{myenumerate}
\end{definition}

There are two cases in which one wants to require stronger
conditions. Firstly, in order to have a correspondence of the Spin
Foam Model with LGT, one seeks a categorical version of the Peter--Weyl
Theorem and of the Haar measure.

The role of the irreducible representations in the Peter--Weyl theory
is now played by the simple objects:

\begin{definition}
Let $\CC$ be a $\C$-linear strict ribbon category.
\begin{myenumerate}
\item
  An object $V$ of $\CC$ is called \emph{simple} if each non-zero
  monomorphism $f\colon U\to V$ is an isomorphism and each non-zero
  epimorphism $f\colon V\to W$ is an isomorphism.
\item
  $\CC$ is called \emph{semi-simple} if each object $V$ of $\CC$ is
  isomorphic to a (finite) direct sum of simple objects.
\item
  $\CC$ is called \emph{finitely} [\emph{countably}]
  \emph{semi-simple} if $\CC$ is semi-simple and if there are only
  finitely [countably] many simple objects up to isomorphism.
\end{myenumerate}
\end{definition}

\begin{corollary}
\label{corr_peterweyl}
Let $\CC$ be a countably semi-simple and admissible ribbon category
such that the unit object $\1$ is simple and such that for each simple
object $J$ we have $\Hom(J,J)\cong\C$. Let $\CC_0$ denote a set
containing one representative per equivalence class of simple objects
of $\CC$ and $V$, $W$ be objects of $\CC$. Then the natural map given
by composition of morphisms,
\begin{equation}
\label{eq_petercateg}
  \bigoplus_{J\in\CC_0}\Hom(V,J)\otimes_\C\Hom(J,W)\to\Hom(V,W),
\end{equation}
is an isomorphism of $\C$-vector spaces.
\end{corollary}

The direct sum in~\eqref{eq_petercateg} plays the role of the
Peter--Weyl decomposition~\eqref{eq_structure_calg} in the categorical
framework.

\begin{remark}
\label{rem_integral}
Under the conditions of Corollary~\ref{corr_peterweyl}, there exists
furthermore for each natural transformation $f_V\colon V\to V$ another
natural transformation ${(Tf)}_V\colon V\to V$ which is defined by the
projection onto the direct summand labelled by $J=\1$
in~\eqref{eq_petercateg},
\begin{equation}
\label{eq_haarcateg}
  T\colon\Hom(V,V)\to\Hom(V,\1)\otimes\Hom(\1,V)\subseteq\Hom(V,V).
\end{equation}
These ${(Tf)}_V$ satisfy, for example, ${(Tf)}_J=0$ for all simple
objects $J$ which are not isomorphic to the unit object $\1$.
\end{remark}

The projection $T$ can be viewed as the translation of the Haar
measure $\int\colon\Calg(G)\to\C$ into the categorical language;
compare~\eqref{eq_haarcateg} with~\eqref{eq_haaralg}.

A detailed explanation of how Corollary~\ref{corr_peterweyl} and
Remark~\ref{rem_integral} are related with Peter--Weyl decomposition
and Haar measure in the Lie group case can be found in~\cite{Oe01}. In
the picture of~\cite{Oe01}, the algebra of representation functions
$\Calg(G)$ of the Lie group $G$ co-acts on the vector spaces dual to
the representations $V_\rho$ of $G$.

A second situation in which one sometimes requires semi-simplicity is
the case when the Spin Foam Model defines a topological invariant,
see, for example~\cite{CrKa97}. Recall, however, that the categories
of finite-dimensional representations of the quantum groups $U_q(\g)$,
$q$ a root of unity, which form important examples, are not
semi-simple~\cite{ChPr94}. I thank R.~Oeckl for pointing out that the
weaker notions of quasi-dominance and dominance (Chapter~XI
of~\cite{Tu94}) can be used to establish a uniform treatment of all
interesting cases.

The problem of the definition of the Spin Foam Model with ribbon
categories which is the subject of the present paper is, however, not
affected by these subtleties since it relies only on
Definition~\ref{def_admissible}. 

\subsubsection{Ribbon diagrams}
\label{sect_ribbondiag}

\begin{figure}[t]
\begin{center}
\input{fig/diag_basic.pstex_t}
\end{center}
\mycaption{fig_diag_basic}{%
  Some basic ribbon diagrams: The identity morphisms $\id_V$ and
  $\id_{V^\ast}$, evaluation $\ev_V$ and co-evaluation $\coev_V$, 
  the braiding $\psi_{V,W}$, the twist $\nu_V$ and their inverses,
  \cf~Definitions~\ref{def_monoidal} to~\ref{def_ribbon}.}
\end{figure}

There exists a very convenient notation for morphisms of a ribbon
category $\CC$ in terms of \emph{ribbon diagrams}
(Figure~\ref{fig_diag_basic}). The diagrams consist of ribbons which
have a white side (normally facing up) and a black side (facing
down). They are directed which is denoted by arrows, and they are
labelled with objects of $\CC$.

The identity morphism $\id_V$ is represented by a ribbon labelled $V$
with the arrow pointing down. The identity morphism $\id_{V^\ast}$ of
the dual object has the same label $V$, but an arrow pointing up. The
diagrams are generally read from top to bottom. Putting diagrams below
each other denotes composition of morphisms while putting them next to
each other denotes the tensor product of
morphisms. Figure~\ref{fig_diag_basic} also shows the natural
isomorphisms $\ev_V$, $\coev_V$ of~\eqref{eq_evcoev}, the braiding
$\psi_{V,W}$ of~\eqref{eq_braiding}, the twist $\nu_V$
of~\eqref{eq_twist} and their inverses, respectively. The unit object
$\1$ is invisible in the diagrams which is justified
by~\eqref{eq_unitobject}, \eqref{eq_trivialbraid}
and~\eqref{eq_trivialtwist}.

\begin{figure}[t]
\begin{center}
\input{fig/diag_dual.pstex_t}
\end{center}
\mycaption{fig_diag_dual}{%
  The conditions~\eqref{eq_straighta} and~\eqref{eq_straightb} on
  evaluation and co-evaluation are depicted in diagrams~(a) and~(b). A
  morphism $f\colon V\to W$ is represented by a coupon as
  in~(c). Diagram (d) shows the definition of the dual morphism
  $f^\ast$ as given by~\eqref{eq_dualmorphism}.} 
\end{figure}

Figure~\ref{fig_diag_dual} shows the conditions~\eqref{eq_straighta}
and~\eqref{eq_straightb} on evaluation and co-evaluation in (a)
and~(b). Morphisms $f\colon V\to W$ are represented by a \emph{coupon}
labelled $f$ with an incoming and an outgoing ribbon as
in~(c). Figure~\ref{fig_diag_dual} also shows the definition of the
dual morphism~\eqref{eq_dualmorphism} in diagram~(d).

\begin{figure}[t]
\begin{center}
\input{fig/diag_ribbon.pstex_t}
\end{center}
\mycaption{fig_diag_ribbon}{%
  The conditions~\eqref{eq_ribbontensor} and~\eqref{eq_ribbondual} on
  the twist $\nu_V$ in diagrammatic notation.}
\end{figure}

\begin{figure}[t]
\begin{center}
\input{fig/diag_rightdual.pstex_t}
\end{center}
\mycaption{fig_diag_rightdual}{%
  Definition of right duality, $\tilde\ev_V$ and $\tilde\coev_V$
  of~\eqref{eq_rightdual}, the quantum trace~\eqref{eq_qtrace} and the
  quantum dimension~\eqref{eq_qdim}.}
\end{figure}

The conditions~\eqref{eq_ribbontensor} and~\eqref{eq_ribbondual} on
the twist $\nu_V$ in a ribbon category are depicted in
Figure~\ref{fig_diag_ribbon}. Figure~\ref{fig_diag_rightdual} shows
the definition of right duals via $\tilde\ev_V$ and $\tilde\coev_V$
of~\eqref{eq_rightdual}, the quantum trace~\eqref{eq_qtrace} and the
quantum dimension~\eqref{eq_qdim}.

The main purpose of the ribbon diagrams presented in this section is
that they have an immediate translation into algebraic language in
terms of morphisms of the ribbon category $\CC$ and at the same time
provide an intuitive way of dealing with the algebraic
manipulations. One can imagine that the ribbons shown in the diagrams
are embedded in $\R^3$. The obvious isotopies then correspond to
relations in $\CC$. This is a direct consequence of the functor
constructed by Reshetikhin and Turaev in~\cite{ReTu90} where more
details can be found. In the following, we present many calculations in
the diagrammatic language. If required, they can be translated at any
stage into the corresponding algebraic expressions.

In the remaining parts of the paper, we employ a simplified notation in
which the ribbons are represented by single directed lines, and it is
understood that their white side always faces up. This is known
as \emph{blackboard framing}. It is particularly convenient here
because it turns out that the relevant diagrams in the following
sections can be drawn without twists.

\subsubsection{Quantum groups and ribbon categories}

The ribbon categories arising in~\cite{ReTu90} are constructed as the
categories of finite-dimensional representations of suitable ribbon
Hopf algebras, see also~\cite{ChPr94,Ma95a,Tu94}.

An alternative picture is developed in~\cite{Oe01}. It is dual to the
former in the sense that it uses the dual Hopf algebra co-acting on
the dual spaces of the representations. It is thus based on coribbon
Hopf algebras and their ribbon category of corepresentations. This
point of view is much closer to the duality transformation for LGT
with Lie groups (see Section~\ref{sect_duality} or~\cite{OePf01})
since the algebra of representation functions $\Calg(G)$ of the gauge
group naturally co-acts on the dual spaces of the representations of
$G$ and can be replaced by a suitable coribbon Hopf
algebra~\cite{Oe01}. 

\subsection{Combinatorial and Simplicial Complexes}

\subsubsection{Triangulations}

For the construction of the Spin Foam Model using ribbon categories, we
need combinatorial complexes and simplicial complexes. Combinatorial
complexes contain the information of which simplices are contained in
the boundary of a given simplex while simplicial complexes also
provide a linear order of the vertices and keep track of all relative
orientations. This terminology follows~\cite{BaWe96}. In order to
construct a Spin Foam Model for ribbon categories, we aim for a
definition of the partition function which takes the relative
orientations into account, but which does not depend on the linear
order of vertices.

For the purpose of the present paper, it is furthermore sufficient to
deal with abstract complexes. The details of how their simplices are
mapped to the given manifold are not discussed here except for a few
restrictions that apply if the complex corresponds to a closed and
oriented manifold.

\begin{definition}
For a given set $\Lambda$ of \emph{vertices}, a \emph{combinatorial
complex} $\Lambda^{(\ast)}$ is a non-empty set of subsets of $\Lambda$,
\begin{equation}
  \emptyset\neq\Lambda^{(\ast)}\subseteq\sym{P}\Lambda,
\end{equation}
such that for each $v\in\Lambda$, $\{v\}\in\Lambda^{(\ast)}$ and for
each set $X\in\Lambda^{(\ast)}$, all its non-empty subsets are also
contained in $\Lambda^{(\ast)}$, \ie\
\begin{equation}
  X\in\Lambda^{(\ast)}\quad\mbox{and}\quad\emptyset\neq Y\subseteq X
  \quad\Longrightarrow\quad
  Y\in\Lambda^{(\ast)}.
\end{equation}
The sets $X\in\Lambda^{(\ast)}$ are called
\emph{simplices}. The subsets $\emptyset\neq Y\subseteq X$ are the
\emph{faces of} $X$. The elements of the set
\begin{equation}
  \Lambda^{(k)} := \{\,X\in\Lambda^{(\ast)}\colon\quad |X|=k+1\,\},\quad
  k\in\N
\end{equation}
are called $k$-simplices. Here $|\cdot|$ denotes the cardinality of a
set. A \emph{combinatorial $k$-complex} is a combinatorial complex for
which $\Lambda^{(j)}=\emptyset$ for all $j>k$. For each $k$-simplex
$X\in\Lambda^{(k)}$, its \emph{boundary} is defined as the collection
of $(k-1)$-simplices,
\begin{equation}
  \del X:=\{\,Y\subseteq X\colon\quad |Y|=k\,\}.
\end{equation}
A combinatorial complex is called \emph{finite} if
$\Lambda^{(\ast)}$ is a finite set.
\end{definition}

\begin{definition}
A \emph{simplicial complex} $(\Lambda^{(\ast)},<)$ is a combinatorial
complex $\Lambda^{(\ast)}$ with a linear order ($<$) of the vertices
$\Lambda$. The $k$-simplices $X\in\Lambda^{(k)}$ can then be
represented by $(k+1)$-tuples $(v_0,v_1,\ldots,v_k)$ of vertices
$v_j\in\Lambda$ in standard order $v_0<v_1<\cdots<v_k$.  In the free
$\Z$-module generated by $\Lambda^{(\ast)}$, the boundary of a
$k$-simplex is given as a sum over the $(k-1)$-simplices in $\del X$,
\begin{equation}
  \del (v_0,v_1,\ldots,v_k) := \sum_{j=0}^k{(-1)}^j\,
    (v_0,v_1,\ldots,\hat v_j,\ldots,v_k),
\end{equation}
where the hat ($\hat\ $) indicates that a symbol is omitted. An
abbreviated notation is
\begin{equation}
  (01\cdots k):=(v_0,v_1,\ldots,v_k).
\end{equation}
\end{definition}

In the following we also use the notation $(v_0,v_1,\ldots,v_k)$ with
arbitrary vertex order. In the simplicial complex this denotes an
oriented $k$-simplex $\sgn\tau\cdot(v_{\tau(0)},\ldots,v_{\tau(k)})$
where the sign depends on the sign of the permutation
$\tau\in\S_{k+1}$ which is required to sort the vertices such that
$v_{\tau(0)}<\cdots< v_{\tau(k)}$.

The triangulations of a compact piecewise-linear $k$-manifold $M$ can
be chosen to have only finitely many simplices. In this case their
combinatorics are described by a finite combinatorial $k$-complex for
which there always exists a linear order of vertices.

For a simplicial $k$-complex which corresponds to the triangulation of
a closed and oriented $k$-manifold $M$, the relative orientation of
each simplex $\sigma$ with respect to $M$ is given, \ie\ whether
$+\sigma$ or $-\sigma$ is isomorphic to a simplex in $M$. Observe
further that in this case each $(k-1)$-simplex is contained in the
boundary of exactly two $k$-simplices: once with positive and once
with negative relative orientation.

\subsubsection{The dual $2$-complex}

In the present paper the Spin Foam Model is defined on a combinatorial
complex $\Lambda^{(\ast)}$. This point of view agrees
with~\cite{BaWe96,CrKa97}, but is dual to the definition given
in~\cite{Ba98}. 

In order to compare the Spin Foam Model on $\Lambda^{(\ast)}$ with
LGT, this LGT has to be formulated on the $2$-complex dual to
$\Lambda^{(\ast)}$. In this section, we define a generalized
notion of $2$-complexes which includes polygons rather than just
triangles and which makes the cyclic ordering of edges around the
polygons explicit. This ordering is necessary to arrange the factors
of the group products which are used in the definition of LGT.

\begin{definition}
\label{def_complexcyclic}
A \emph{finite generalized $2$-complex with cyclic structure}
$(V,E,F)$ consists of finite sets $V$ (\emph{vertices}), $E$
(\emph{edges}) and $F$ (\emph{polygons}) together with maps
\begin{mathletters}
\label{eq_twocomplex}
\begin{alignat}{2}
  \del_+&\colon E\to V,&\qquad&\mbox{(end point of an edge)}\\
  \del_-&\colon E\to V,&\qquad&\mbox{(starting point of an edge)}\\
  N     &\colon F\to\N,&\qquad&\mbox{(number of edges in the boundary
                        of a polygon)}\\
  \del_j&\colon F\to E,&\qquad&\mbox{(the $j$-th edge in the boundary)}\\
  \epsilon_j&\colon F\to\{-1,+1\},&\qquad&\mbox{(its orientation)}
\end{alignat}%
\end{mathletters}%
such that
\begin{mathletters}
\label{eq_boundary}
\begin{gather}
  \del_{-\epsilon_jf}\del_jf = \del_{\epsilon_{j+1}f}\del_{j+1}f,\qquad
    1\leq j\leq N(f)-1,\\
  \del_{-\epsilon_{N(f)}f}\del_{N(f)}f = \del_{\epsilon_1f}\del_1f,
\end{gather}%
\end{mathletters}%
for all $f\in F$.
\end{definition}

\begin{figure}[t]
\begin{center}
\input{fig/complex_boundary.pstex_t}
\end{center}
\mycaption{fig_boundary}{%
  The maps $\del_j$ and $\epsilon_j$ and the
  conditions~\eqref{eq_boundary}. Here $N(f)=3$, $\epsilon_1f=+1$,
  $\epsilon_2f=+1$ and $\epsilon_3f=(-1)$.} 
\end{figure}

The conditions~\eqref{eq_boundary} state that the edges in the
boundary of a polygon $f\in F$ are in cyclic ordering from $\del_1f$
to $\del_{N(f)}f$ where one encounters the edges with a relative
orientation given by $-\epsilon_jf$, see
Figure~\ref{fig_boundary}. Observe that~\eqref{eq_boundary}
can be used to generalize the condition $\del\circ\del=0$ to the
situation where the edges are labelled with non-commutative
variables. 

Given a finite simplicial $k$-complex $\Lambda^{(\ast)}$, one can
construct the dual $2$-complex $(V,E,F)$ in the standard way: The dual
vertices are just the $k$-simplices, $V:=\Lambda^{(k)}$. The dual
edges are the $(k-1)$-simplices, $E:=\Lambda^{(k-1)}$, and the dual
polygons are given by the $(k-2)$-simplices,
$F:=\Lambda^{(k-2)}$. Observe that the $(k-2)$-simplices
$\Lambda^{(k-2)}$ are in general contained in the boundaries of more
than three $(k-1)$-simplices which implies that the polygons $F$ have
in general more than three edges. The maps $\del_j$, $\epsilon_j$,
\etc\ of~\eqref{eq_twocomplex} can be constructed inductively from the
boundary relation of the simplicial complex.

In the calculations of the next section, the following abbreviations
are convenient: For a given edge $e\in E$, the sets
\begin{mathletters}
\begin{eqnarray}
e_+ &:=& \{f\in F\colon\quad e=\del_j f,\quad\epsilon_j f=(+1)\quad
           \mbox{for some $j$,}\quad 1\leq j\leq N(f)\},\\
e_- &:=& \{f\in F\colon\quad e=\del_j f,\quad\epsilon_j f=(-1)\quad
           \mbox{for some $j$,}\quad 1\leq j\leq N(f)\},
\end{eqnarray}%
\end{mathletters}%
contain all polygons that have the edge $e$ in their boundary with
positive ($+$) or negative ($-$) orientation. For a given polygon
$f\in F$, the set
\begin{equation}
  f_0 := \{v\in V\colon\quad v=\delta_-\delta_jf\quad
           \mbox{for some $j$,}\quad 1\leq j\leq N(f)\},
\end{equation}
denotes all vertices that are contained in the boundary of the polygon
$f$. Finally, the sets 
\begin{mathletters}
\begin{eqnarray}
  f_+ &:=& \{e\in E\colon\quad e=\del_jf,\quad\epsilon_jf=(+1)\quad
             \mbox{for some $j$,}\quad 1\leq j\leq N(f)\},\\
  f_- &:=& \{e\in E\colon\quad e=\del_jf,\quad\epsilon_jf=(-1)\quad
             \mbox{for some $j$,}\quad 1\leq j\leq N(f)\},
\end{eqnarray}%
\end{mathletters}%
denote all edges in the boundary of the polygon $f$ with positive ($+$)
or negative ($-$) orientation.

%
\section{The duality transformation}
%
\label{sect_duality}

In this section, we recall the duality transformation relating LGT for
Lie groups on the dual generalized $2$-complex $(V,E,F)$ with a spin
foam model. This transformation was carried out in~\cite{OePf01} on a
hypercubic lattice and is formulated here for generic $2$-complexes.

The calculation is presented entirely in terms of the $2$-complex
$(V,E,F)$ and does not refer to the simplicial complex
$\Lambda^{(\ast)}$. Its relation with $\Lambda^{(\ast)}$ will be
discussed in the following section. The calculation is furthermore
valid in arbitrary dimension $d\geq 2$.

\begin{definition}
\label{def_partition}
Let $G$ be a compact Lie group (or a finite group). The partition
function of LGT on the finite generalized $2$-complex $(V,E,F)$ with
cyclic structure is defined by
\begin{equation}
\label{eq_partfunc}
  Z = \Bigl(\prod_{e\in E}\int_G\,dg_e\Bigr)\,
      \prod_{f\in F}w(g_f),\qquad
      g_f:=g_{\del_1f}^{\epsilon_1f}\cdots g_{\del_{N(f)}f}^{\epsilon_{N(f)}f}
\end{equation}
Here $\int_G$ denotes the normalized Haar measure on $G$, $w\colon
G\to\R$ is the Boltzmann weight~\eqref{eq_boltzmann}, and $g_f$ is the
cyclicly ordered product of the group elements attached to the edges
in the boundary of the polygon $f\in F$.
\end{definition}

\begin{remark}
\begin{myenumerate}
\item
  Observe that even though this definition explicitly refers to the
  cyclic structure, the value of $Z$ is actually independent of
  it. The starting point for the cyclic numbering of edges in the
  boundary of a polygon does not matter because the Boltzmann weight
  is given by a class function and thus invariant under cyclic
  permutation of the factors of $g_f$. Reversal of the orientation is
  also a symmetry because it replaces $g_f$ by $g_f^{-1}$ which yields
  the complex conjugate of the class function, but this function is
  real.
\item
  Let $h\colon V\to G,v\mapsto h_v$ associate a group element to each
  vertex. The weight $w(g_f)$ in~\eqref{eq_partfunc} is invariant
  under the \emph{local gauge transformations},
\begin{equation}
  g_e\mapsto \alignidx{h_{\del_+e}\cdot g_e\cdot h_{\del_-e}^{-1}}.
\end{equation}
  In order to prove this invariance, one has to make use of the
  conditions~\eqref{eq_boundary}.
\end{myenumerate}
\end{remark}

The first step of the duality transformation is to insert the
character expansion~\eqref{eq_charexp} of the Boltzmann weight
into~\eqref{eq_partfunc},
\begin{equation}
  Z = \Bigl(\prod_{e\in E}\int_G\,dg_e\Bigr)\,
      \prod_{f\in F}\sum_{\rho_f\in\Irrep}\hat w_{\rho_f}\,
      \sum_{n_f=1}^{\dim V_{\rho_f}}t^{(\rho_f)}_{n_fn_f}(g_f).
\end{equation}
The trace of the character is responsible for summations over one
index $n_f$ per polygon $f\in F$. The application of coproduct and
antipode (eq.~\eqref{eq_matrixcopro} and~\eqref{eq_matrixanti}) to the
product $g_f$ (eq.~\eqref{eq_partfunc}) yields further vector index
summations. In total there is one summation per polygon and per vertex
of that polygon. These summation variables are denoted by $n(f,v)$
where $f\in F$ and $v\in f_0$,
\begin{eqnarray}
  Z &=& \Bigl(\prod_{e\in E}\int_G\,dg_e\Bigr)\,
      \prod_{f\in F}\sum_{\rho_f\in\Irrep}\hat w_{\rho_f}\,
      \sum_{n(f,\del_-\del_1f)=1}^{\dim V_{\rho_f}}\cdots
      \sum_{n(f,\del_-\del_{N(f)}f)=1}^{\dim V_{\rho_f}}\nn\\
  &&\quad  t^{(\rho_f)}_{n(f,\del_+\del_1f),n(f,\del_-\del_1f)}(g_{\del_1f}^{\epsilon_1f})
             \cdots
 t^{(\rho_f)}_{n(f,\del_+\del_{N(f)}f),n(f,\del_-\del_{N(f)}f)}(g_{\del_{N(f)}f}^{\epsilon_{N(f)}f}).
\end{eqnarray}
Recall that the conditions~\eqref{eq_boundary} of the $2$-complex apply
here. The above expression can now be reorganized, moving all
summations to the left,
\begin{eqnarray}
  Z &=& \Bigl(\prod_{e\in E}\int_G\,dg_e\Bigr)\,
        \Bigl(\prod_{f\in F}\sum_{\rho_f\in\Irrep}\Bigr)\,
        \Bigl(\prod_{f\in F}\hat w_{\rho_f}\Bigr)\,
        \Bigl(\prod_{f\in F}\prod_{v\in f_0}\sum_{n(f,v)=1}^{\dim V_{\rho_f}}\Bigr)\nn\\
  &&\quad \prod_{f\in F}
    \Bigl(\prod_{e\in f_+}t^{(\rho_f)}_{n(f,\del_+e),n(f,\del_-e)}(g_e)\Bigr)\,
    \Bigl(\prod_{e\in f_-}t^{(\rho^\ast_f)}_{n(f,\del_-e),n(f,\del_+e)}(g_e)\Bigr).
\end{eqnarray}
Here the notation
\begin{equation}
  \Bigl(\prod_{f\in F}\sum_{\rho_f\in\Irrep}\Bigr):=
  \sum_{\rho_f\in\Irrep}\cdots\sum_{\rho_f\in\Irrep}
\end{equation}
denotes one summation per polygon $f\in F$. Sorting the product of
representation functions by edge rather than by polygon amounts to
just a slight change in the enumeration of polygons and edges,
\begin{eqnarray}
  Z &=& \Bigl(\prod_{f\in F}\sum_{\rho_f\in\Irrep}\Bigr)\,
        \Bigl(\prod_{f\in F}\hat w_{\rho_f}\Bigr)\,
        \Bigl(\prod_{f\in F}\prod_{v\in f_0}\sum_{n(f,v)=1}^{\dim V_{\rho_f}}\Bigr)\,
        \prod_{e\in E}\int_G\,dg_e\Biggl[\Biggr.\nn\\
  &&\quad\Biggl.\prod_{e\in E}
    \Bigl(\prod_{f\in e_+}t^{(\rho_f)}_{n(f,\del_+e),n(f,\del_-e)}(g_e)\Bigr)\,
    \Bigl(\prod_{f\in e_-}t^{(\rho^\ast_f)}_{n(f,\del_-e),n(f,\del_+e)}(g_e)\Bigr)\Biggr].
\end{eqnarray}
The integrals can now be evaluated using the formula~\eqref{eq_haaralg},
\begin{eqnarray}
\label{eq_intsolved}
 && \int_G\,dg_e[\cdots] =\nn\\
 && \sum_{P^{(e)}\in\sym{P}_e}
    P^{(e)\overbrace{n(f,\del_+e),\ldots}^{f\in e_+}
          \overbrace{n(f,\del_-e),\ldots}^{f\in e_-}}\,
    P^{(e)}_{\underbrace{n(f,\del_-e),\ldots}_{f\in e_+}
             \underbrace{n(f,\del_+e),\ldots}_{f\in e_-}},
\end{eqnarray}
where $\sym{P}_e$ denotes a basis of orthogonal $G$-invariant
projectors onto the trivial components in the complete decomposition
of
\begin{equation}
\label{eq_decomp}
  \bigl(\bigotimes_{f\in e_+}\rho_f\bigr)\otimes
  \bigl(\bigotimes_{f\in e_-}\rho_f^\ast\bigr).
\end{equation}
The curly brackets in~\eqref{eq_intsolved} indicate that there is one
index $n(f,\del_+e)$ for each $f\in e_+$ \etc. Finally, the sums
over projectors are moved to the left of the expression,
\begin{eqnarray}
  Z &=& \Bigl(\prod_{f\in F}\sum_{\rho_f\in\Irrep}\Bigr)\,
        \Bigl(\prod_{e\in E}\sum_{P^{(e)}\in\sym{P}_e}\Bigr)\,
        \Bigl(\prod_{f\in F}\hat w_{\rho_f}\Bigr)\,
        \Bigl(\prod_{f\in F}\prod_{v\in f_0}\sum_{n(f,v)=1}^{\dim V_{\rho_f}}\Bigr)\nn\\
  &&\quad\Biggl.\prod_{e\in E}
    P^{(e)\overbrace{n(f,\del_+e),\ldots}^{f\in e_+}
          \overbrace{n(f,\del_-e),\ldots}^{f\in e_-}}\,
    P^{(e)}_{\underbrace{n(f,\del_-e),\ldots}_{f\in e_+}
             \underbrace{n(f,\del_+e),\ldots}_{f\in e_-}}.
\end{eqnarray}
This formula can now be reorganized and yields the final result:

\begin{theorem}
\label{thm_dualpart}
The partition function~\eqref{eq_partfunc} of LGT on the finite
generalized $2$-complex $(V,E,F)$ with cyclic structure is equal to
the expression
\begin{equation}
\label{eq_dual_partition}
  Z = \Bigl(\prod_{f\in F}\sum_{\rho_f\in\Irrep}\Bigr)\,
      \Bigl(\prod_{e\in E}\sum_{P^{(e)}\in\sym{P}_e}\Bigr)\,
      \Bigl(\prod_{f\in F}\hat w_{\rho_f}\Bigr)\,
      \Bigl(\prod_{v\in V}C(v)\Bigr)
\end{equation}
Here $\sym{P}_e$ denotes a basis of orthogonal $G$-invariant
projectors onto the trivial components in the complete decomposition
of~\eqref{eq_decomp}. The weights per polygon $\hat w_{\rho_f}$ are
the coefficients of the character expansion~\eqref{eq_charexp} of the
original Boltzmann weight. The weights per vertex $C(v)$ are given by
a trace involving representations and projectors in the neighbourhood
of the vertex $v\in V$,
\begin{equation}
\label{eq_cv}
  C(v) = \Bigl(\prod_{\ontop{f\in F\colon}{v\in f_0}}\sum_{n_f=1}^{\dim V_{\rho_f}}\Bigr)\,
         \Bigl(\prod_{\ontop{e\in E\colon}{v=\del_+e}}
           P^{(e)\overbrace{n_fn_f\ldots}^{f\in e_+}
                 \overbrace{n_fn_f\ldots}^{f\in e_-}}\Bigr)\,
         \Bigl(\prod_{\ontop{e\in E\colon}{v=\del_-e}}
           P^{(e)}_{\underbrace{n_fn_f\ldots}_{f\in e_+}
                    \underbrace{n_fn_f\ldots}_{f\in e_-}}\Bigr).
\end{equation}
Here the range $f\in F\colon v\in f_0$ of the first product refers to
all polygons $f\in F$ that contain the vertex $v$ in their boundary.
\end{theorem}

\begin{figure}[t]
\begin{center}
\input{fig/duality_partition.pstex_t}
\end{center}
\mycaption{fig_duality_partition}{%
  The neighbourhood of a vertex $v\in V$ on the dual $2$-complex in
  the three-dimensional case. The dotted lines denote the four edges
  attached to the vertex. Diagram (a) shows the weight $C(v)$ per
  vertex $v\in V$ occurring in the Spin Foam Model where
  the full dots denote projectors $P^{(e)}$, and the solid lines the
  representations $V_\rho$. Diagram (b) visualizes the weight $\tilde
  C(v)$ in the spin network expectation value. Here $Q^{(v)}$ is the
  morphism attached to $v$, and the dashed lines denote the
  representations $\tau_e$.} 
\end{figure}

\begin{remark}
\begin{myenumerate}
\item
  The projectors onto the trivial representations,
\begin{equation}
  P^{(e)}\colon\bigl(\bigotimes_{f\in e_+}\rho_f\bigr)\otimes
         \bigl(\bigotimes_{f\in e_-}\rho_f^\ast\bigr)\to\C,
\end{equation}
  can be replaced via the isomorphisms $\Hom(V\otimes
  W^\ast,\C)\cong\Hom(V,W)$ by representation morphisms
\begin{equation}
  \phi^{(e)}\colon\bigotimes_{f\in e_+}\rho_f\to
            \bigotimes_{f\in e_-}\rho_f.
\end{equation}
  The partition function then contains a sum over a basis of the space
  of representation morphisms for each edge $e\in E$,
\begin{equation}
  \Hom(\bigotimes_{f\in e_+}\rho_f,\bigotimes_{f\in e_-}\rho_f).
\end{equation}
\item
  The expression $C(v)$ is a trace in the category of finite
  dimensional representations $\RCat G$,
  \cf~Figure~\ref{fig_duality_partition}(a). Observe that all vector
  indices $n_f$ are contracted. The complexity of the $C(v)$
  depends on the number of edges which contain $v\in V$ in their
  boundary. In order to generalize this Spin Foam Model to ribbon
  categories, $C(v)$ has to be replaced by a quantum trace. The main
  motivation for formulating LGT on the $2$-complex dual to a
  triangulation is that it is now guaranteed that in dimension $4$
  there are always precisely $5$ edges which contain $v$. Without this
  restriction, the generalization of $C(v)$ to the ribbon case would
  be much harder.
\item
  Observe that for $G=\SU(2)$ in 3 dimensions, the $C(v)$ are
  essentially the $6j$-symbols of $\SU(2)$.
\end{myenumerate}
\end{remark}

The generic observables of LGT that have non-vanishing expectation
values under the path integral are \emph{spin networks}, the
generalization of Wilson loops to the non-Abelian case.

\begin{definition}
\label{def_spinnet}
Let $G$ be a compact Lie group (or a finite group), $(V,E,F)$ be a
finite generalized $2$-complex with cyclic structure and $Z$ denote
the partition function of $LGT$ of Definition~\ref{def_partition}. Let
$\tau\colon E\to\Irrep$ assign a unitary finite-dimensional
irreducible representation $\tau_e$ to each edge $e\in E$ and for each
vertex $v\in V$, let
\begin{equation}
  Q^{(v)}\colon\bigotimes_{\ontop{e\in E\colon}{v=\del_+e}}\tau_e\to
               \bigotimes_{\ontop{e\in E\colon}{v=\del_-e}}\tau_e
\end{equation}
denote a representation morphism. The \emph{spin network} labelled by
$\tau_e$ and $Q^{(v)}$ associates to each configuration $E\to
G,e\mapsto g_e$ the value
\begin{equation}
\label{eq_spinnet}
  W(\tau,Q) := \frac{1}{Z}
    \Bigl(\prod_{e\in E}\sum_{k_e,\ell_e=1}^{\dim V_{\tau_e}}\Bigr)\,
    \Bigl(\prod_{e\in E}t_{k_e\ell_e}^{(\tau_e)}(g_e)\Bigr)\,
    \Bigl(\prod_{v\in V}
      Q^{(v)}_{\underbrace{k_e\ldots}_{\ontop{e\in E\colon}{v=\del_+e}}
               \underbrace{\ell_e\ldots}_{\ontop{e\in E\colon}{v=\del_-e}}}\Bigr).
\end{equation}
\end{definition}

For more details on this definition and for the proof of the following
result, we refer the reader to~\cite{OePf01}.

\begin{theorem}
Let $G$ be a compact Lie group (or a finite group) and $(V,E,F)$ be a
finite generalized $2$-complex with cyclic structure. Let $\tau_e$ and
$Q^{(v)}$ define a spin network as in
Definition~\ref{def_spinnet}. The expectation value of the spin
network,
\begin{equation}
  \left<W(\tau,Q)\right> = 
    \Bigl(\prod_{e\in E}\int_G\,dg_e\Bigr)\,\Bigl[
      W(\tau,Q)\prod_{f\in F}w(g_f)\Bigr],
\end{equation}
is equal to 
\begin{eqnarray}
\label{eq_dualspinnet}
  \left<W(\tau,Q)\right> &=& \frac{1}{Z}\,
    \Bigl(\prod_{f\in F}\sum_{\rho_f\in\Irrep}\Bigr)\,
    \Bigl(\prod_{e\in E}\sum_{P^{(e)}\in\tilde{\sym{P}}_e}\Bigr)\,
    \Bigl(\prod_{f\in F}\hat w_{\rho_f}\Bigr)\nn\\
  &&\quad \prod_{v\in V}\Bigl[
    \Bigl(\prod_{\ontop{e\in E}{v=\del_+e}}\sum_{k_e=1}^{\dim V_{\tau_e}}\Bigr)\,
    \Bigl(\prod_{\ontop{e\in E}{v=\del_-e}}\sum_{\ell_e=1}^{\dim V_{\tau_e}}\Bigr)\,
      \tilde C(v)\cdot 
      Q^{(v)}_{\underbrace{k_e\ldots}_{\ontop{e\in E\colon}{v=\del_+e}}
               \underbrace{\ell_e\ldots}_{\ontop{e\in E\colon}{v=\del_-e}}}\Bigr].
\end{eqnarray}
Here $\tilde{\sym{P}}_e$ is a basis of orthogonal $G$-invariant
projectors onto the trivial components in the complete decomposition
of 
\begin{equation}
  \bigl(\bigotimes_{f\in e_+}\rho_f\bigr)\otimes
  \bigl(\bigotimes_{f\in e_-}\rho_f^\ast\bigr)\otimes\tau_e.
\end{equation}
The weights per polygon $\hat w_{\rho_f}$ are the coefficients of the
character expansion~\eqref{eq_charexp} of the original Boltzmann
weight. The weights per vertex $\tilde C(v)$ are given by the trace
\begin{equation}
  \tilde C(v) = \Bigl(\prod_{\ontop{f\in F\colon}{v\in f_0}}\sum_{n_f=1}^{\dim V_{\rho_f}}\Bigr)\,
         \Bigl(\prod_{\ontop{e\in E\colon}{v=\del_+e}}
           P^{(e)\overbrace{n_fn_f\ldots}^{f\in e_+}
                 \overbrace{n_fn_f\ldots}^{f\in e_-}
                 k_e}\Bigr)\,
         \Bigl(\prod_{\ontop{e\in E\colon}{v=\del_-e}}
           P^{(e)}_{\underbrace{n_fn_f\ldots}_{f\in e_+}
                    \underbrace{n_fn_f\ldots}_{f\in e_-}
                    \ell_e}\Bigr).
\end{equation}
\end{theorem}

\begin{remark}
\begin{myenumerate}
\item
  The above theorem is an example for an explicit calculation how the
  spin foams that are the configurations in the partition function
  couple to the spin network
  $W(\tau,Q)$. Figure~\ref{fig_duality_partition}(b) visualizes the
  trace which gives the weights per vertex $\tilde C(v)$.
\item
  If the set of edges for which the representations $\tau_e$ are
  non-trivial, forms a closed loop, then $W(\tau,Q)$ is non-zero only
  if the non-trivial $\tau_e$ are all isomorphic. The morphisms
  $Q^{(v)}$ are then unique up to normalization. In this case
  $W(\tau,Q)$ describes a Wilson loop.
\end{myenumerate}
\end{remark}

%
\section{The Spin Foam Model for Ribbon Categories}
%
\label{sect_ribbonlgt}

\begin{table}
\begin{center}
\begin{tabular}{l|l|l|l}
  triangulation & dual $2$-complex & colouring     & weights       \\
\hline
  $4$-simplex   & vertex           & ---           & $C(v)$        \\
  tetrahedron   & edge             & morphism      & ---           \\
  triangle      & polygon          & simple object & $\hat w_\rho$ \\
  edge          & ---              & ---           & ---           \\
  vertex        & ---              & ---           & ---           \\
\end{tabular}
\mycaption{tab_dualcomplex}{%
  The partition function of the Spin Foam Model dual to
  LGT~\eqref{eq_dual_partition} is a sum over all colourings where the
  summands contain certain weights. Here colourings and weights are given
  for LGT living on the $2$-complex dual to the triangulation.}
\end{center}
\end{table}

In Section~\ref{sect_duality}, the Spin Foam Model dual to LGT was
derived for the case in which the gauge group $G$ is a Lie
group. If LGT is defined on the $2$-complex dual to the triangulation,
the partition function of the Spin Foam Model consists of a sum over
all labellings of triangles with irreducible representations (simple
objects) and of all tetrahedra with invariant projectors
(representation morphisms), see Table~\ref{tab_dualcomplex}.

This Spin Foam Model shall be generalized to a ribbon category $\CC$
which replaces the category of representations $\RCat G$ of the gauge
group $G$. The partition function will contain the sum over all
colourings of triangles with simple objects explicitly while the sum
over all colourings of tetrahedra with morphisms will be implemented
as a trace over suitable state spaces.

The definition of the Spin Foam Model is formulated in a first step
for a given simplicial $4$-complex. The definition thus refers
explicitly to the linear order of vertices. In a second step we will
prove that it does not depend on that order and that it is thus
well-defined for any combinatorial complex that corresponds to the
triangulation of a closed and oriented piecewise-linear $4$-manifold.

\subsection{Definition of the partition function}
\label{sect_ribbondef}

First we define the colourings which will be explicitly summed over in the
partition function.

\begin{definition}
\label{def_colouring}
Let $\Lambda^{(\ast)}$ denote a simplicial complex, $\CC$ be an
admissible ribbon category (Definition~\ref{def_admissible}) and
$\CC_0$ a set of colours.
\begin{myenumerate}
\item
  A \emph{colouring} $V\colon\Lambda^{(2)}\to\CC_0$ associates an
  object $V(v_0,v_1,v_2)\in\CC_0$ to each triangle
  $(v_0,v_1,v_2)\in\Lambda^{(2)}$ with standard vertex order
  $v_0<v_1<v_2$.
\item
  For any permutation $\sigma\in\S_3$ (acting on $\{0,1,2\}$) define
\begin{equation}
\label{eq_colourperm}
  V(v_{\sigma(0)},v_{\sigma(1)},v_{\sigma(2)}):=\left\{
    \begin{matrix}
      V(v_0,v_1,v_2),        & \mbox{if}\quad\sgn\sigma = 1,\\
      {V(v_0,v_1,v_2)}^\ast, & \mbox{if}\quad\sgn\sigma = -1.
    \end{matrix}\right.
\end{equation}
\end{myenumerate}
For given vertices $v_0,v_1,v_2\in\Lambda$, we use the abbreviated
notation 
\begin{equation}
  V_{012} := V(v_0,v_1,v_2),\qquad V_{021} := V(v_0,v_2,v_1),
\end{equation}
and so on, for example, $\alignidx{V_{021}=V_{012}^\ast}$.
\end{definition}

Recall that ${(V^\ast)}^\ast\cong V$ is isomorphic in $\CC$, but in
general not equal. The above definition therefore describes an action
of the symmetric group $\sym{S}_3$ only up to isomorphism.

The state spaces are defined in the next step. A trace over these
spaces will yield the summation over colourings of the tetrahedra with
morphisms.

\begin{figure}[t]
\begin{center}
\input{fig/morphism.pstex_t}
\end{center}
\mycaption{fig_morphism}{%
  (a) The coupons denoting morphisms $\phi_{0123}\in H_{0123}$ and their duals
  $\phi_{0123}^\ast$ \eqref{eq_statespace_std} as well as (b) morphisms of
  the dual state spaces, $\overline\phi_{0123}\in H_{0123}^\ast$ and
  their duals $\overline\phi_{0123}^\ast$
  \eqref{eq_statespace_dual}. Diagram (c) shows the
  pairing~\eqref{eq_pairhom}. All ribbons are drawn in blackboard
  framing.}
\end{figure}

\begin{definition}
\label{def_statespace}
Let $V\colon\Lambda^{(2)}\to\CC_0$ denote a colouring. The \emph{state
space} associated with a tetrahedron $(v_0,v_1,v_2,v_3)$ with
arbitrary vertex order is defined by
\begin{mathletters}
\begin{equation}
  H^{(V)}(v_0,v_1,v_2,v_3) := 
    \Hom(V(v_1,v_2,v_3)\otimes V(v_0,v_1,v_3),V(v_0,v_1,v_2)\otimes V(v_0,v_2,v_3)).
\end{equation}%
The dual state space is then given up to isomorphism by the pairing~\eqref{eq_pairing},
\begin{equation}
\label{eq_statespace_dual}
  {H^{(V)}(v_0,v_1,v_2,v_3)}^\ast :=
    \Hom({V(v_0,v_1,v_3)}^\ast\otimes{V(v_1,v_2,v_3)}^\ast,
         {V(v_0,v_2,v_3)}^\ast\otimes{V(v_0,v_1,v_2)}^\ast).
\end{equation}%
\end{mathletters}%
The following abbreviated notation is used,
\begin{mathletters}
\begin{eqnarray}
\label{eq_statespace_std}
  H_{0123} &=& \Hom(V_{123}\otimes V_{013},V_{012}\otimes V_{023}),\\
  H_{0123}^\ast &=&
    \Hom(V_{013}^\ast\otimes V_{123}^\ast,V_{023}^\ast\otimes V_{012}^\ast),
\end{eqnarray}%
\end{mathletters}%
so that the pairing~\eqref{eq_pairing} reads in this case
\begin{equation}
\label{eq_pairhom}
  {\left<\cdot,\cdot\right>}_{0123}\colon H_{0123}^\ast\otimes H_{0123}\to\C,
  (\overline\phi_{0123},\psi_{0123})\mapsto
    {\left<\overline\phi_{0123},\psi_{0123}\right>}_{0123}.
\end{equation}
\end{definition}

The ribbon diagrams corresponding to a morphism $\phi_{0123}\in
H_{0123}$ and its dual $\phi_{0123}^\ast$ are depicted in
Figure~\ref{fig_morphism}(a). The morphism of the dual state space
$\overline\phi_{0123}\in H_{0123}^\ast$ and its dual
$\overline\phi_{0123}^\ast$ are represented diagrammatically as in
Figure~\ref{fig_morphism}(b). Dual morphisms are denoted by a star
(${}^\ast$) whereas we indicate by a bar ($\overline{\ \ \vbox to
1ex{\vfill}}$) that a morphism belongs to a dual state
space. Figure~\ref{fig_morphism}(c) shows the
pairing~\eqref{eq_pairhom} for morphisms $\overline\phi_{0123}\in
H_{0123}^\ast$ and $\psi_{0123}\in H_{0123}$. In
Figure~\ref{fig_morphism} and in the following we use blackboard
framing (see Section~\ref{sect_ribbondiag}).

\begin{remark}
\begin{myenumerate}
\item
  Definition~\ref{def_statespace} applies to any order of
  vertices. In particular, the definition of $H_{0123}$ involves 
  $V(v_0,v_1,v_2)$ \etc\ as given by~\eqref{eq_colourperm}.
\item
  The definition~\eqref{eq_statespace_dual} implements a special
  choice of isomorphism between $H_{0123}$ and $H_{0123}^\ast$ via the
  pairing~\eqref{eq_pairhom}. This choice is used consistently in the
  following.
\end{myenumerate}
\end{remark}

\begin{figure}[t]
\begin{center}
\input{fig/sigma_01.pstex_t}
\mycaption{fig_sigma_01}{%
  The definition of the morphisms
  $\phi_{1023}:={\tau_0^\ast}^{-1}(\overline\phi_{0123})\in H_{1023}$
  and $\overline\phi_{1023}:=\tau_0(\phi_{0123})\in H^\ast_{1023}$ for
  given $\overline\phi_{0123}\in H^\ast_{0123}$ and  
  $\phi_{0123}\in H_{0123}$, see~\eqref{eq_sigma_01}.}
\end{center}
\end{figure}

\begin{figure}[th]
\begin{center}
\input{fig/sigma_12.pstex_t}
\mycaption{fig_sigma_12}{%
  The definition of the morphisms
  $\phi_{0213}:={\tau_1^\ast}^{-1}(\overline\phi_{0123})\in H_{0213}$
  and $\overline\phi_{0213}:=\tau_1(\phi_{0123})\in H^\ast_{0213}$ for
  given $\overline\phi_{0123}\in H^\ast_{0123}$ and 
  $\phi_{0123}\in H_{0123}$, see~\eqref{eq_sigma_12}.}
\end{center}
\end{figure}

\begin{figure}[th]
\begin{center}
\input{fig/sigma_23.pstex_t}
\mycaption{fig_sigma_23}{%
  The definition of the morphisms 
  $\phi_{0132}:={\tau_2^\ast}^{-1}(\overline\phi_{0123})\in H_{0132}$
  and $\overline\phi_{0132}:=\tau_2(\phi_{0123})\in H^\ast_{0132}$ for
  given $\overline\phi_{0123}\in H^\ast_{0123}$ and
  $\phi_{0123}\in H_{0123}$, see~\eqref{eq_sigma_23}.}
\end{center}
\end{figure}

The state spaces of Definition~\ref{def_statespace} for different
vertex order are related by linear isomorphisms which are defined
in the next step.

\begin{definition}
\label{def_snaction}
Let $H_{0123}$ and $H_{0123}^\ast$ denote the state space and its dual
for a tetrahedron $(v_0,v_1,v_2,v_3)$. The linear maps
\begin{mathletters}
\begin{alignat}{2}
\label{eq_sigma_01}
  \tau_0&\colon H_{0123}\to H_{1023}^\ast,\qquad&
    {\tau_0^\ast}^{-1}&\colon H_{0123}^\ast\to H_{1023},\\
\label{eq_sigma_12}
  \tau_1&\colon H_{0123}\to H_{0213}^\ast,\qquad&
    {\tau_1^\ast}^{-1}&\colon H_{0123}^\ast\to H_{0213},\\
\label{eq_sigma_23}
  \tau_2&\colon H_{0123}\to H_{0132}^\ast,\qquad&
    {\tau_2^\ast}^{-1}&\colon H_{0123}^\ast\to H_{0132},
\end{alignat}%
\end{mathletters}%
are defined by the diagrams in Figure~\ref{fig_sigma_01} to
Figure~\ref{fig_sigma_23}. 
\end{definition}

Note that $\tau_j$ exchanges the $j$-th and the $(j+1)$-th vertex of
the four arguments of $H(v_0,v_1,v_2,v_3)$, counting from zero. This
need not be the vertices with number $j$ and $j+1$, for example,
\begin{equation}
  \tau_0\colon H_{1234}\to H_{2134}^\ast,\qquad
  \tau_1\colon H_{0214}\to H_{0124}^\ast.
\end{equation}

\begin{figure}[t]
\begin{center}
\input{fig/sigma_inverse.pstex_t}
\mycaption{fig_sigma_inverse}{%
  Diagrammatic proof of the identity 
  ${\tau_1^\ast}^{-1}\circ\tau_1=\id_{H_{0123}}$ in
  Lemma~\ref{lemma_inverse}. Here $\phi_{0123}\in H_{0123}$.} 
\end{center}
\end{figure}

\begin{lemma}
\label{lemma_inverse}
Let $\tau_j$, $0\leq j\leq 2$, denote the linear maps of
Definition~\ref{def_snaction}. 
\begin{myenumerate}
\item
  The $\tau_j$ satisfy ${\tau_j^\ast}^{-1}\circ\tau_j=\id$ and
  $\tau_j\circ{\tau_j^\ast}^{-1}=\id$. In particular, the $\tau_j$ and
  ${\tau_j^\ast}^{-1}$ form linear isomorphisms.
\item
  The $\tau_j$ satisfy
  ${\left<{\tau_j^\ast}^{-1}(\psi_{0123}),\tau_j(\overline\phi_{0123})\right>}_{1023}
  ={\left<\overline\phi_{0123},\psi_{0123}\right>}_{0123}$ for all
  $\overline\phi_{0123}\in H_{0123}^\ast$ and $\psi_{0123}\in
  H_{0123}$ which motivates the notation ${\tau_j^\ast}^{-1}$. 
\end{myenumerate}
\end{lemma}

\begin{proof}
\begin{myenumerate}
\item
  The relations ${\tau_j^\ast}^{-1}\circ\tau_j=\id$ can be verified
  diagrammatically using the identities that hold in ribbon
  categories. Figure~\ref{fig_sigma_inverse} shows the calculation for
  ${\tau_1^\ast}^{-1}\circ\tau_1=\id_{H_{0123}}$. The other cases are
  analogous. 
\item
  This claim can also be verified diagrammatically. It is essentially
  a consequence of the fact that the maps $\tau_j$ on the dual state
  spaces in Figure~\ref{fig_sigma_01} to Figure~\ref{fig_sigma_23} are
  given by the mirror images of the maps on the original state spaces.
\end{myenumerate}
\end{proof}

\begin{figure}[t]
\begin{minipage}[t]{7.4cm}
\begin{center}
\input{fig/vertex_std.pstex_t}
\end{center}
\end{minipage}
\begin{minipage}[t]{7.4cm}
\begin{center}
\input{fig/vertex_opp.pstex_t}
\end{center}
\end{minipage}
\mycaption{fig_vertex}{%
  (a) The quantum trace~\eqref{eq_trace_std} defining
  the $4$-simplex map for a $4$-simplex with positive relative
  orientation in $M$. (b) The quantum trace~\eqref{eq_trace_opp}
  for negative relative orientation. The morphisms
  are denoted by $\phi_{jk\ell m}\in H_{jk\ell m}$ and 
  $\overline\phi_{jk\ell m}\in H_{jk\ell m}^\ast$ and represented by
  the coupons of Figure~\ref{fig_morphism}.}
\end{figure}

\begin{remark}
In analogy with the three-dimensional case, one could conjecture that
the $\tau_j$ generate an action of the symmetric group $\sym{S}_4$ on
some collection of state spaces. This is not the case. Only in the
final step we will have an action of the symmetric group when it is
proved that the partition function is well-defined.
\end{remark}

At this point, the colourings $V_{jk\ell}$ and the spaces $H_{jk\ell
m}$ are defined for a generic vertex order. The summation over all
coulourings of tetrahedra with morphisms which is part of the
partition function, will be implemented as a trace. This trace is over
the tensor product of maps $Z^{(V)}_{01234}$ for all $4$-simplices
$(v_0,\ldots,v_4)\in\Lambda^{(4)}$. These building blocks
$Z^{(V)}_{01234}$ are defined first.

\begin{definition}
\label{def_simplexmap}
Let $V\colon\Lambda^{(2)}\to\CC_0$ be a colouring and the state spaces
for the tetrahedra be given by Definition~\ref{def_statespace}.
\begin{myenumerate}
\item
  For any $4$-simplex $(v_0,\ldots,v_4)$ whose relative orientation in
  the manifold $M$ is positive, define the $4$-\emph{simplex map}
\begin{mathletters}
\begin{equation}
  Z^{(V),(+)}_{01234}\colon 
    H_{0234}\otimes H_{0124}\to H_{1234}\otimes H_{0134}\otimes H_{0123},
\end{equation}
  to be the linear map that is related by the pairing~\eqref{eq_pairing}
  to the quantum trace
\begin{equation}
\label{eq_trace_std}
  Z^{\prime(V),(+)}_{01234}\colon H_{1234}^\ast\otimes H_{0234}\otimes H_{0134}^\ast\otimes
    H_{0124}\otimes H_{0123}^\ast\to\C,
\end{equation}%
\end{mathletters}%
  which is depicted in Figure~\ref{fig_vertex}(a). 
\item
  For any $4$-simplex with negative relative orientation in $M$, the
  $4$-\emph{simplex map}
\begin{mathletters}
\begin{equation}
  Z^{(V),(-)}_{01234}\colon 
    H_{1234}\otimes H_{0134}\otimes H_{0123}\to H_{0234}\otimes H_{0124},
\end{equation}
  is defined by the quantum trace
\begin{equation}
\label{eq_trace_opp}
  Z^{\prime(V),(-)}_{01234}\colon H_{1234}\otimes H_{0234}^\ast\otimes H_{0134}\otimes
    H_{0124}^\ast\otimes H_{0123}\to\C, 
\end{equation}%
\end{mathletters}%
  which is depicted in Figure~\ref{fig_vertex}(b). 
\end{myenumerate}
\end{definition}

\begin{remark}
\begin{myenumerate}
\item
  The assignment of the $H_{jk\ell m}$ to domain or codomain and the
  assignment of duality stars (${}^\ast$) in the above definitions is
  according to the orientation of the tetrahedra in the
  boundary of the $4$-simplex,
\begin{equation}
  \del (01234) = (1234) - (0234) + (0134) - (0124) + (0123).
\end{equation}
\item
  Observe that Figure~\ref{fig_vertex}(b) is the mirror image of (a)
  with all arrows reversed. This is different from the quantum trace
  of the dual morphism which would also replace the over-crossing by
  an under-crossing.
\end{myenumerate}
\end{remark}

In order to obtain a summation over a basis of each state space
$H_{jk\ell m}$, the partition function is defined as a trace over the
tensor product of all $4$-simplex maps.

\begin{definition}
\label{def_prelimpart}
Let $\Lambda^{(\ast)}$ be a finite combinatorial $4$-complex
corresponding to a triangulation of a closed oriented piecewise-linear
$4$-manifold $M$. Choose a fixed linear order of the vertices of
$\Lambda$. Let $V\colon\Lambda^{(2)}\to\CC_0$ be a colouring and let
the $4$-simplex maps $Z^{(V),(\pm)}_{jk\ell mn}$ be given by
Definition~\ref{def_simplexmap}.

The \emph{partition function per colouring} is defined as
\begin{equation}
\label{eq_prelimpart}
  Z^{(V)} := \tr_{\sym{H}} \biggl[P\circ
    \Bigl(\bigotimes_{(v_0,\ldots,v_4)\in\Lambda^{(4)}} 
    Z^{(V),(\epsilon_{01234})}_{01234}\Bigr)\biggr].
\end{equation}
Here $\epsilon_{01234}\in\{+1,-1\}$ denotes the relative orientation
of the $4$-simplex $(v_0,\ldots,v_4)\in\Lambda^{(4)}$ in $M$. Since
every tetrahedron occurs precisely twice in the boundary of a
$4$-simplex, once with positive and once with negative relative
orientation, both domain and codomain of the tensor product
over the $Z^{(V),(\pm)}_{01234}$ are permutations of the tensor
factors of
\begin{equation}
  \sym{H}:=\bigotimes_{(v_0,v_1,v_2,v_3)\in\Lambda^{(3)}} H^{(V)}(v_0,v_1,v_2,v_3).
\end{equation}
The permutation operator $P$ in~\eqref{eq_prelimpart} is the unique
permutation which sorts the tensor factors of the codomain such that
their ordering agrees with the ordering of factors in the domain.
\end{definition}

\begin{remark}
  The trace over the tensor product $\sym{H}$ in the above definition
  essentially contains the quantum traces~\eqref{eq_trace_std}
  or~\eqref{eq_trace_opp} for all $4$-simplices plus an additional
  summation over bases of all state spaces. In the partition function
  the traces generalize the weights $C(v)$, \cf\
  Table~\ref{tab_dualcomplex}
  and~\eqref{eq_cv}. Figure~\ref{fig_vertex} is the four-dimensional
  analogue of Figure~\ref{fig_duality_partition}(a) with a particular
  choice of over- and under-crossings.
\end{remark}

\begin{definition}
\label{def_partribbon}
Let $\Lambda^{(\ast)}$ be a finite combinatorial $4$-complex
corresponding to a triangulation of a closed oriented piecewise-linear
$4$-manifold $M$. Choose a fixed linear order of the vertices of
$\Lambda$. For each colouring $V\colon\Lambda^{(2)}\to\CC_0$, let the
partition function per colouring, $Z^{(V)}$, be given by
Definition~\ref{def_prelimpart}. The \emph{partition function} is
defined as
\begin{equation}
\label{eq_partribbon}
  Z := \sum_{V\colon\Lambda^{(2)}\to\CC_0}\biggl(
    \prod_{(v_0,v_1,v_2)\in\Lambda^{(2)}}\hat
    w(v_0,v_1,v_2)(V_{012})\biggr)\, Z^{(V)}. 
\end{equation}
The weights $\hat w(v_0,v_1,v_2)\colon\CC_0\to\R$ assign a real number
to the object associated with the triangle
$(v_0,v_1,v_2)\in\Lambda^{(2)}$ and are required to satisfy the
\emph{reality condition},
\begin{equation}
\label{eq_reality}
  \hat w(v_0,v_1,v_2)(V^\ast)=\hat w(v_0,v_1,v_2)(V),
\end{equation}
and to be functions on equivalence classes of isomorphic objects, \ie
\begin{equation}
  \hat w(v_0,v_1,v_2)(V) = \hat w(v_0,v_1,v_2)(\tilde V)
    \qquad\mbox{if}\quad V\cong\tilde V.
\end{equation}
\end{definition}

Section~\ref{sect_welldef} is devoted to proving that this definition
is actually independent of the linear order of vertices and of the
choice of colours $\CC_0$ up to isomorphism. The partition function is
therefore well-defined for a combinatorial complex that corresponds to
the triangulation of a closed and oriented manifold. In
Section~\ref{sect_special}, we discuss some relevant special cases in
more detail that are covered by~\eqref{eq_partribbon}, in particular
the relation with the standard formulation of LGT for Lie groups and
the Crane--Yetter state sum. There, we also comment on the convergence
of~\eqref{eq_partribbon} if $\CC_0$ is not a finite set.

\subsection{Properties of the partition function}
\label{sect_welldef}

First we show that the partition function~\eqref{eq_partribbon} does
not depend on the choice of colours $\CC_0$ up to isomorphism.

\begin{theorem}
\label{thm_simple}
Let $V\colon\Lambda^{(2)}\to\CC_0$ denote a colouring, and for each
triangle $(v_0,v_1,v_2)\in\Lambda^{(2)}$, $v_0<v_1<v_2$, let
\begin{equation}
  \Phi(v_0,v_1,v_2)\colon V(v_0,v_1,v_2)\to\tilde V(v_0,v_1,v_2)
\end{equation}
be an isomorphism in $\CC$ for some object $\tilde V(v_0,v_1,v_2)$.
Then the partition functions per colouring~\eqref{eq_prelimpart} for
$V$ and $\tilde V$ agree,
\begin{equation}
  Z^{(V)} = Z^{(\tilde V)}.
\end{equation}
\end{theorem}

\begin{proof}
Using the standard abbreviations, the given isomorphisms are of the
form $\Phi_{012}\colon V_{012}\to\tilde V_{012}$ for all triangles
$(v_0,v_1,v_2)$ with standard vertex order $v_0<v_1<v_2$. For any
permutation $\sigma\in\S_{3}$ we define isomorphisms 
$\Phi_{\sigma(0)\sigma(1)\sigma(2)}\colon 
  V_{\sigma(0)\sigma(1)\sigma(2)}\to \tilde
V_{\sigma(0)\sigma(1)\sigma(2)}$ by
\begin{equation}
  \Phi_{\sigma(0)\sigma(1)\sigma(2)}:=\left\{
    \begin{matrix}
      \Phi_{012}\colon V_{012}\to\tilde V_{012},
      &\mbox{if}\quad\sgn\sigma = 1,\\
      {\Phi^\ast}^{-1}_{012}\colon V^\ast_{012}\to {\tilde V}^\ast_{012},
      &\mbox{if}\quad\sgn\sigma = -1.
    \end{matrix}\right.
\end{equation}
Observe that this assignment is compatible with
Definition~\ref{def_colouring}. These definitions provide us with
isomorphisms $\Phi_{012}\colon V_{012}\to\tilde V_{012}$ and with
their dual maps $\Phi_{012}^\ast\colon{\tilde V}^\ast_{012}\to
V_{012}^\ast$ for all triangles $(v_0,v_1,v_2)$ with arbitrary vertex
order. 

Furthermore, there are induced linear isomorphisms of the state
spaces,
\begin{mathletters}
\label{eq_simpleiso}
\begin{eqnarray}
  \Phi_{0123}\colon\Hom(V_{123}\otimes V_{013},V_{012}\otimes V_{023})
    &\to&\Hom(\tilde V_{123}\otimes\tilde V_{013},\tilde V_{012}\otimes\tilde V_{023}),\\
  \phi_{0123}&\mapsto&
    (\Phi_{012}\otimes\Phi_{023})\circ\phi_{0123}\circ(\Phi^{-1}_{123}\otimes\Phi^{-1}_{013}),\nn
\end{eqnarray}
and
\begin{eqnarray}
  \Phi^\ast_{0123}\colon
    \Hom({\tilde V}^\ast_{013}\otimes{\tilde V}^\ast_{123},
         {\tilde V}^\ast_{023}\otimes{\tilde V}^\ast_{012})
    &\to&\Hom(V^\ast_{013}\otimes V^\ast_{123},V_{023}^\ast\otimes V^\ast_{012}),\\
  \overline\phi_{0123}&\mapsto&
    (\Phi^\ast_{023}\otimes\Phi^\ast_{012})\circ\overline\phi_{0123}
    \circ({\Phi^\ast}^{-1}_{013}\otimes{\Phi^\ast}^{-1}_{123}).\nn
\end{eqnarray}%
\end{mathletters}%
A convenient abbreviated notation for these maps is $\Phi_{0123}\colon
H_{0123}\to\tilde H_{0123}$, $\Phi^\ast_{0123}\colon{\tilde
H}^\ast_{0123}\to H^\ast_{0123}$ writing $\tilde H_{0123}:=\Hom(\tilde
V_{123}\otimes\tilde V_{013},\tilde V_{012}\otimes\tilde V_{023})$
\etc. Now the following diagram for the traces
$Z^{\prime(V),(+)}_{01234}$ commutes:
\begin{center}
\setlength{\unitlength}{7mm}
\begin{picture}(14,6)
  \put(1,5){\makebox(0,0){$H_{1234}^\ast$}}
  \put(2,5){\makebox(0,0){$\otimes$}}
  \put(3,5){\makebox(0,0){$H_{0234}$}}
  \put(4,5){\makebox(0,0){$\otimes$}}
  \put(5,5){\makebox(0,0){$H_{0134}^\ast$}}
  \put(6,5){\makebox(0,0){$\otimes$}}
  \put(7,5){\makebox(0,0){$H_{0124}$}}
  \put(8,5){\makebox(0,0){$\otimes$}}
  \put(9,5){\makebox(0,0){$H_{0123}^\ast$}}

  \put(1,1){\makebox(0,0){${\tilde H}_{1234}^\ast$}}
  \put(2,1){\makebox(0,0){$\otimes$}}
  \put(3,1){\makebox(0,0){${\tilde H}_{0234}$}}
  \put(4,1){\makebox(0,0){$\otimes$}}
  \put(5,1){\makebox(0,0){${\tilde H}_{0134}^\ast$}}
  \put(6,1){\makebox(0,0){$\otimes$}}
  \put(7,1){\makebox(0,0){${\tilde H}_{0124}$}}
  \put(8,1){\makebox(0,0){$\otimes$}}
  \put(9,1){\makebox(0,0){${\tilde H}_{0123}^\ast$}}

  \put(1,4.4){\vector(0,-1){3}}
  \put(1.9,3){\makebox(0,0){${\Phi^\ast}^{-1}_{1234}$}}
  \put(3,4.4){\vector(0,-1){3}}
  \put(3.8,3){\makebox(0,0){$\Phi_{0234}$}}
  \put(5,4.4){\vector(0,-1){3}}
  \put(5.9,3){\makebox(0,0){${\Phi^\ast}^{-1}_{0134}$}}
  \put(7,4.4){\vector(0,-1){3}}
  \put(7.8,3){\makebox(0,0){$\Phi_{0124}$}}
  \put(9,4.4){\vector(0,-1){3}}
  \put(9.9,3){\makebox(0,0){${\Phi^\ast}^{-1}_{0123}$}}

  \put(10,4.5){\vector(2,-1){2}}
  \put(11.6,4.7){\makebox(0,0){$Z_{01234}^{\prime(V),(+)}$}}
  \put(10,1.5){\vector(2,1){2}}
  \put(11.6,1.3){\makebox(0,0){$Z_{01234}^{\prime(\tilde V),(+)}$}}
  \put(12.5,3){\makebox(0,0){$\C$}}
\end{picture}
\end{center}
To see this, imagine Figure~\ref{fig_vertex}(a) drawn for maps
$\tilde\phi_{jk\ell m}\in\tilde H_{jk\ell m}$ \etc\ and insert the
definitions of the linear isomorphisms $\Phi_{jk\ell m}$
of~\eqref{eq_simpleiso}. Then the isomorphisms in $\CC$,
$\Phi_{jk\ell}\colon V_{jk\ell}\to\tilde V_{jk\ell}$, appear twice in
each ribbon in a way such that they cancel. 

Let ${\left<\cdot,\cdot\right>}^{\tilde{}}\colon 
{\tilde H}^\ast_{0123}\otimes\tilde H_{0123}\to\C$ denote the
pairing~\eqref{eq_pairhom} applied to the state spaces which use the
colouring $\tilde V$. We find
\begin{equation}
  {\left<{\Phi^\ast}^{-1}_{0123}(\overline{\phi}_{0123}),
  \Phi_{0123}(\psi_{0123})\right>}^{\tilde{}}
  =\left<\overline{\phi}_{0123},\psi_{0123}\right>,
\end{equation}
for all $\overline{\phi}_{0123}\in H_{0123}^\ast$ and $\psi_{0123}\in
H_{0123}$. As a consequence the following diagram involving the
$4$-simplex maps themselves also commutes:
\begin{center}
\setlength{\unitlength}{7mm}
\begin{picture}(13,6)
  \put(1,5){\makebox(0,0){$H_{0234}$}}
  \put(2,5){\makebox(0,0){$\otimes$}}
  \put(3,5){\makebox(0,0){$H_{0124}$}}

  \put(4,5){\vector(1,0){2}}

  \put(7,5){\makebox(0,0){$H_{1234}$}}
  \put(8,5){\makebox(0,0){$\otimes$}}
  \put(9,5){\makebox(0,0){$H_{0134}$}}
  \put(10,5){\makebox(0,0){$\otimes$}}
  \put(11,5){\makebox(0,0){$H_{0123}$}}

  \put(1,1){\makebox(0,0){${\tilde H}_{0234}$}}
  \put(2,1){\makebox(0,0){$\otimes$}}
  \put(3,1){\makebox(0,0){${\tilde H}_{0124}$}}

  \put(4,1){\vector(1,0){2}}

  \put(7,1){\makebox(0,0){${\tilde H}_{1234}$}}
  \put(8,1){\makebox(0,0){$\otimes$}}
  \put(9,1){\makebox(0,0){${\tilde H}_{0134}$}}
  \put(10,1){\makebox(0,0){$\otimes$}}
  \put(11,1){\makebox(0,0){${\tilde H}_{0123}$}}

  \put(1,4.4){\vector(0,-1){3}}
  \put(1.8,3){\makebox(0,0){$\Phi_{0234}$}}
  \put(3,4.4){\vector(0,-1){3}}
  \put(3.8,3){\makebox(0,0){$\Phi_{0124}$}}
  \put(7,4.4){\vector(0,-1){3}}
  \put(7.8,3){\makebox(0,0){$\Phi_{1234}$}}
  \put(9,4.4){\vector(0,-1){3}}
  \put(9.8,3){\makebox(0,0){$\Phi_{0134}$}}
  \put(11,4.4){\vector(0,-1){3}}
  \put(11.8,3){\makebox(0,0){$\Phi_{0123}$}}

  \put(5,5.5){\makebox(0,0){$Z_{01234}^{(V),(+)}$}}
  \put(5,1.5){\makebox(0,0){$Z_{01234}^{(\tilde V),(+)}$}}
\end{picture}
\end{center}
Analogous diagrams are available for $Z^{(V),(-)}_{01234}$ and
Figure~\ref{fig_vertex}(b) in the case of opposite orientation.

Finally, each tetrahedron occurs precisely twice in the boundaries of
$4$-simplices, once with positive and once with negative relative
orientation. Therefore the tensor product of all $4$-simplex maps
in~\eqref{eq_prelimpart} is conjugated by a linear isomorphism $\Phi$
which can be obtained from a tensor product of the $\Phi_{jk\ell m}$,
\begin{equation}
\label{eq_conjugiso}
  P\circ\Bigl(\bigotimes_{\sigma\in\Lambda^{(4)}}
    Z_\sigma^{(V),(\epsilon_\sigma)}\Bigr)
  = \Phi\,\Bigl[ P\circ\Bigl(\bigotimes_{\sigma\in\Lambda^{(4)}} 
    Z_{\sigma}^{(\tilde V),(\epsilon_\sigma)}\Bigr)\Bigr]\,\Phi^{-1}.
\end{equation}
Since $Z^{(V)}$ is the trace of~\eqref{eq_conjugiso}, it agrees with
$Z^{(\tilde V)}$.
\end{proof}

\begin{corollary}
The partition function~\eqref{eq_partribbon} does not depend on the
choice of colours $\CC_0$ up to isomorphism.
\end{corollary}

\begin{proof}
Consider another set of colours $\tilde\CC_0$ such that each colouring
$V\colon\Lambda^{(2)}\to\CC_0$ induces a colouring $\tilde
V\colon\Lambda^{(2)}\to\tilde\CC_0$ for which $V_{012}\cong\tilde
V_{012}$ are isomorphic in $\CC$ for all triangles
$(v_0,v_1,v_2)\in\Lambda^{(2)}$. The partition
function~\eqref{eq_partribbon} defined using $\CC_0$ agrees with that
one defined using $\tilde\CC_0$ because the weights satisfy $\hat
w(v_0,v_1,v_2)(V_{012})=\hat w(v_0,v_1,v_2)(\tilde V_{012})$ and
because $Z^{(V)}=Z^{(\tilde V)}$ according to
Theorem~\ref{thm_simple}.
\end{proof}

In order to prove the independence of the partition
function~\eqref{eq_partribbon} of the linear order of vertices, a
generic $4$-simplex $(01234)$ is considered. It is proved that any
permutation of its vertices which results in different $4$-simplex
maps $H_{jk\ell m}$ according to Definition~\ref{def_statespace} and
Definition~\ref{def_simplexmap}, does not change the partition
function~\eqref{eq_partribbon}.

This statement is verified for the four elementary transpositions of
$\S_5$ (acting on the vertices $\{0,1,2,3,4\}$). The following lemmas
prepare the proof. They establish diagrammatical isotopies which
permute the coupons in Figure~\ref{fig_vertex}(a) in order to reach a
configuration similar to Figure~\ref{fig_vertex}(b). Recall that the
orientation of the $4$-simplex changes if an odd permutation is
applied to its vertices.

\begin{lemma}
\label{lemma_perm01}
For any colouring $V\colon\Lambda^{(2)}\to\CC_0$ and morphisms
$\phi_{jk\ell m}\in H_{jk\ell m}$ and $\overline\phi_{jk\ell m}\in
H_{jk\ell m}^\ast$, the quantum trace in Figure~\ref{fig_vertex}(a) is
equal to the quantum trace in Figure~\ref{fig_vertex_apply_01}.
\end{lemma}

\begin{proof}
The calculation is described in diagrammatic language and can be
translated into equalities for morphisms of the ribbon category $\CC$
as described in Section~\ref{sect_ribbon}. First, a number of coupons
are moved around in the plane without twisting or braiding any
ribbons: Move the coupon $\phi_{0124}$ to the left and place it above
$\overline\phi_{1234}$, then move $\overline\phi_{0134}$ down and to
the right and place it below and right of $\phi_{0234}$. Move
$\overline\phi_{0123}$ to the right and place it below
$\overline\phi_{0123}$ and below and left of
$\overline\phi_{0134}$. Rotate the coupon $\overline\phi_{0123}$ by
$360$ degrees in order to place its ribbons as depicted in
Figure~\ref{fig_vertex_apply_01}. Finally, lift the ribbon labelled
$V_{014}$ out of the plane, move it across the entire diagram, and
place it as shown in Figure~\ref{fig_vertex_apply_01}.
\end{proof}

\begin{lemma}
\label{lemma_perm12}
For any colouring $V\colon\Lambda^{(2)}\to\CC_0$ and morphisms
$\phi_{jk\ell m}\in H_{jk\ell m}$ and $\overline\phi_{jk\ell m}\in
H_{jk\ell m}^\ast$, the quantum trace in Figure~\ref{fig_vertex}(a) is
equal to the quantum trace in Figure~\ref{fig_vertex_apply_12}.
\end{lemma}

\begin{proof}
The proof is again explained diagrammatically: Start with
Figure~\ref{fig_vertex}(a). Lift the ribbon $V_{012}$ out of the plane
and move it across the coupon $\phi_{1234}$ so that $V_{012}$ now
over-crosses $V_{123}$, $V_{134}$ and $V_{124}$ rather than
$V_{234}$. Then the coupons can be moved around in the plane without
introducing twists or braidings such that the configuration in
Figure~\ref{fig_vertex_apply_12} is obtained.
\end{proof}

There exist two more lemmas that deal with the elementary
transpositions $(23)$ and $(34)$ as well as four lemmas dealing with
the case of opposite relative orientation. They are not stated
explicitly here since they are very similar and completely analogous
to prove.

The results of the preceding lemmas, Figure~\ref{fig_vertex_apply_01}
and Figure~\ref{fig_vertex_apply_12}, are furthermore related to the
quantum trace of Figure~\ref{fig_vertex}(b) for a $4$-simplex with a
different order of vertices. This is stated in the following lemmas.

\begin{lemma}
\label{lemma_commutative01}
Let $\tau=(01)$ and consider the $4$-simplex $(01234)$. Let
$V\colon\Lambda^{(2)}\to\CC_0$ denote a colouring. Then there exists
another colouring $\tilde V$ with isomorphic objects for each
triangle, ${\tilde V}_{012}\cong V_{012}$, such that the following
diagram commutes:
\begin{center}
\setlength{\unitlength}{7mm}
\begin{picture}(13,6)
  \put(1,5){\makebox(0,0){$H_{0234}$}}
  \put(2,5){\makebox(0,0){$\otimes$}}
  \put(3,5){\makebox(0,0){$H_{0124}$}}

  \put(4,5){\vector(1,0){2}}

  \put(7,5){\makebox(0,0){$H_{1234}$}}
  \put(8,5){\makebox(0,0){$\otimes$}}
  \put(9,5){\makebox(0,0){$H_{0134}$}}
  \put(10,5){\makebox(0,0){$\otimes$}}
  \put(11,5){\makebox(0,0){$H_{0123}$}}

  \put(1,1){\makebox(0,0){$H_{0234}$}}
  \put(2,1){\makebox(0,0){$\otimes$}}
  \put(3,1){\makebox(0,0){$H^\ast_{1024}$}}

  \put(4,1){\vector(1,0){2}}

  \put(7,1){\makebox(0,0){$H_{1234}$}}
  \put(8,1){\makebox(0,0){$\otimes$}}
  \put(9,1){\makebox(0,0){$H^\ast_{1034}$}}
  \put(10,1){\makebox(0,0){$\otimes$}}
  \put(11,1){\makebox(0,0){$H^\ast_{1023}$}}

  \put(1,4.4){\vector(0,-1){3}}
  \put(1.4,3){\makebox(0,0){$\id$}}
  \put(3,4.4){\vector(0,-1){3}}
  \put(3.4,3){\makebox(0,0){$\tau_0$}}
  \put(7,4.4){\vector(0,-1){3}}
  \put(7.4,3){\makebox(0,0){$\id$}}
  \put(9,4.4){\vector(0,-1){3}}
  \put(9.4,3){\makebox(0,0){$\tau_0$}}
  \put(11,4.4){\vector(0,-1){3}}
  \put(11.4,3){\makebox(0,0){$\tau_0$}}
  \put(5,5.5){\makebox(0,0){$Z_{01234}^{(V),(+)}$}}
\end{picture}
\end{center}
Here $Z^{(V),(+)}_{01234}$ is the $4$-simplex map of
Definition~\ref{def_simplexmap}, the maps $\tau_0$ are given in
Definition~\ref{def_snaction}, and the bottom horizontal map is
determined, using the pairing~\eqref{eq_pairhom}, by the $4$-simplex
map
\begin{equation}
  Z^{(\tilde V)(-)}_{10234}\colon H_{0234}\otimes H_{1034}\otimes H_{1023}
  \to H_{1234}\otimes H_{1024}.
\end{equation}
\end{lemma}

\begin{proof}
Consider Figure~\ref{fig_vertex_apply_01} whose quantum trace agrees
with $Z^{\prime(V),(+)}_{01234}$ of Figure~\ref{fig_vertex}(a)
according to Lemma~\ref{lemma_perm01}. The linear isomorphisms
$\tau_0$ of Definition~\ref{def_snaction} can now be used to replace
the dashed boxes of Figure~\ref{fig_vertex_apply_01} by morphisms of
the state spaces $H_{jk\ell m}$ with a different vertex order.
The result is very similar to the trace $Z^{\prime(V)(-)}_{10234}$ of
Figure~\ref{fig_vertex}(b) for the $4$-simplex $(10234)$.

Observe, however, that in Figure~\ref{fig_vertex}(b) the arrows of the
ribbons corresponding to the triangles $(012)$, $(013)$ and $(014)$
are reversed compared with Figure~\ref{fig_vertex_apply_01}. We can
reverse these arrows in Figure~\ref{fig_vertex_apply_01} if we label
them instead by $V^\ast_{012}$, $V^\ast_{013}$ and $V^\ast_{014}$,
respectively.

Consider the triangle $(012)$. If $(v_0,v_1,v_2)$ is an even
permutation of the standard vertex order, then
Figure~\ref{fig_vertex_apply_01} contains $V_{012}=V$ for some object
$V\in\CC_0$, \ie\ upon reversal of the arrows this label changes to
$V^\ast_{012}=V_{102}$. This is the same label as the label arising in
$Z^{\prime(V)(-)}_{10234}$. 

If, however, $(v_0,v_1,v_2)$ is an odd permutation of the standard
vertex order, then Figure~\ref{fig_vertex_apply_01} contains
$V_{012}=V^\ast$ for some object $V\in\CC_0$, \ie\ upon arrow reversal
this becomes $V_{012}^\ast={(V^\ast)}^\ast$. This is in general not
identical, but still isomorphic to $V$ which arises in
$Z^{\prime(V)(-)}_{10234}$ in this case. This is the reason why the
present lemma holds only for a colouring $\tilde V$ with isomorphic
objects at all triangles. 

Let $(w_0,w_1,w_2)$ denote any triangle in standard vertex order,
$w_0<w_1<w_2$. Define the colouring $\tilde V$ by
\begin{equation}
  {\tilde V}_{\sigma(0)\sigma(1)\sigma(2)}:=\left\{
    \begin{matrix}
      {({V_{012}^\ast})}^\ast,
      &\mbox{if}\quad\sgn\sigma = 1,\\
      V_{012}^\ast,
      &\mbox{if}\quad\sgn\sigma = -1,
    \end{matrix}\right.
\end{equation}
if
$\{w_0,w_1,w_2\}\in\{\{v_0,v_1,v_2\},\{v_0,v_1,v_3\},\{v_0,v_1,v_4\}\}$
and by ${\tilde
V}_{\sigma(0)\sigma(1)\sigma(2)}:=V_{\sigma(0)\sigma(1)\sigma(2)}$ for
the other triangles. Then the quantum trace of
Figure~\ref{fig_vertex_apply_01} with arrows $(012)$, $(013)$ and
$(014)$ reversed, agrees with the trace of Figure~\ref{fig_vertex}(b)
for the colouring $\tilde V$. The following diagram therefore commutes:
\begin{center}
\setlength{\unitlength}{7mm}
\begin{picture}(15,5)
  \put(1,4){\makebox(0,0){$H_{1234}^\ast$}}
  \put(2,4){\makebox(0,0){$\otimes$}}
  \put(3,4){\makebox(0,0){$H_{0234}$}}
  \put(4,4){\makebox(0,0){$\otimes$}}
  \put(5,4){\makebox(0,0){$H_{0134}^\ast$}}
  \put(6,4){\makebox(0,0){$\otimes$}}
  \put(7,4){\makebox(0,0){$H_{0124}$}}
  \put(8,4){\makebox(0,0){$\otimes$}}
  \put(9,4){\makebox(0,0){$H_{0123}^\ast$}}

  \put(1,1){\makebox(0,0){$H_{0234}$}}
  \put(2,1){\makebox(0,0){$\otimes$}}
  \put(3,1){\makebox(0,0){$H_{1234}^\ast$}}
  \put(4,1){\makebox(0,0){$\otimes$}}
  \put(5,1){\makebox(0,0){$H_{1034}$}}
  \put(6,1){\makebox(0,0){$\otimes$}}
  \put(7,1){\makebox(0,0){$H_{1024}^\ast$}}
  \put(8,1){\makebox(0,0){$\otimes$}}
  \put(9,1){\makebox(0,0){$H_{1023}$}}

  \put(1,3.5){\vector(1,-1){2}}
  \put(0.8,3){\makebox(0,0){$\id$}}
  \put(3,3.5){\vector(-1,-1){2}}
  \put(3.2,3){\makebox(0,0){$\id$}}
  \put(5,3.5){\vector(0,-1){2}}
  \put(5.8,2.5){\makebox(0,0){${\tau_0^\ast}^{-1}$}}
  \put(7,3.5){\vector(0,-1){2}}
  \put(7.4,2.5){\makebox(0,0){$\tau_0$}}
  \put(9,3.5){\vector(0,-1){2}}
  \put(9.8,2.5){\makebox(0,0){${\tau_0^\ast}^{-1}$}}

  \put(10,4){\vector(2,-1){2}}
  \put(11.6,4.2){\makebox(0,0){$Z_{01234}^{\prime(V),(+)}$}}
  \put(10,1){\vector(2,1){2}}
  \put(11.6,0.8){\makebox(0,0){$Z_{10234}^{\prime(\tilde V),(-)}$}}
  \put(12.5,2.5){\makebox(0,0){$\C$}}
\end{picture}
\end{center}
Using Lemma~\ref{lemma_inverse}, this implies the commutativity of the
diagram claimed in the present lemma.
\end{proof}

\begin{lemma}
\label{lemma_commutative12}
Let $\tau=(12)$ and $V\colon\Lambda^{(2)}\to\CC_0$ be a
colouring. Then there exists another colouring $\tilde V$ with
isomorphic objects for each triangle such that the following diagram
commutes:
\begin{center}
\setlength{\unitlength}{7mm}
\begin{picture}(13,6)
  \put(1,5){\makebox(0,0){$H_{0234}$}}
  \put(2,5){\makebox(0,0){$\otimes$}}
  \put(3,5){\makebox(0,0){$H_{0124}$}}

  \put(4,5){\vector(1,0){2}}

  \put(7,5){\makebox(0,0){$H_{1234}$}}
  \put(8,5){\makebox(0,0){$\otimes$}}
  \put(9,5){\makebox(0,0){$H_{0134}$}}
  \put(10,5){\makebox(0,0){$\otimes$}}
  \put(11,5){\makebox(0,0){$H_{0123}$}}

  \put(1,1){\makebox(0,0){$H_{0234}$}}
  \put(2,1){\makebox(0,0){$\otimes$}}
  \put(3,1){\makebox(0,0){$H^\ast_{0214}$}}

  \put(4,1){\vector(1,0){2}}

  \put(7,1){\makebox(0,0){$H^\ast_{2134}$}}
  \put(8,1){\makebox(0,0){$\otimes$}}
  \put(9,1){\makebox(0,0){$H_{0134}$}}
  \put(10,1){\makebox(0,0){$\otimes$}}
  \put(11,1){\makebox(0,0){$H^\ast_{0213}$}}

  \put(1,4.4){\vector(0,-1){3}}
  \put(1.4,3){\makebox(0,0){$\id$}}
  \put(3,4.4){\vector(0,-1){3}}
  \put(3.4,3){\makebox(0,0){$\tau_1$}}
  \put(7,4.4){\vector(0,-1){3}}
  \put(7.4,3){\makebox(0,0){$\tau_0$}}
  \put(9,4.4){\vector(0,-1){3}}
  \put(9.4,3){\makebox(0,0){$\id$}}
  \put(11,4.4){\vector(0,-1){3}}
  \put(11.4,3){\makebox(0,0){$\tau_1$}}
  \put(5,5.5){\makebox(0,0){$Z_{01234}^{(V),(+)}$}}
\end{picture}
\end{center}
Here the $\tau_j$ are the isomorphisms given in
Definition~\ref{def_snaction}, and the bottom horizontal map is
determined, using the pairing~\eqref{eq_pairhom}, by the $4$-simplex
map
\begin{equation}
  Z^{(\tilde V),(-)}_{02134}\colon 
    H_{2134}\otimes H_{0234}\otimes H_{0213}\to
    H_{0134}\otimes H_{0214}.
\end{equation}
\end{lemma}

\begin{proof}
Consider Figure~\ref{fig_vertex_apply_12} whose quantum trace agrees
with $Z^{\prime(V),(+)}_{01234}$ of Figure~\ref{fig_vertex}(a)
according to Lemma~\ref{lemma_perm12}. The linear isomorphisms
$\tau_0$ and $\tau_1$ of Definition~\ref{def_snaction} can now be used
to replace the dashed boxes of Figure~\ref{fig_vertex_apply_12} by
morphisms of the state spaces $H_{jk\ell m}$ with a different vertex
order. The result is the trace $Z_{02134}^{\prime(\tilde V),(-)}$
of Figure~\ref{fig_vertex}(b) for the $4$-simplex $(02134)$ up to the
choice of isomorphic objects for the triangles $(012)$, $(123)$ and
$(124)$. These isomorphisms arise from double dualization as in
Lemma~\ref{lemma_commutative01}. 

The following diagram therefore commutes:
\begin{center}
\setlength{\unitlength}{7mm}
\begin{picture}(15,5)
  \put(1,4){\makebox(0,0){$H_{1234}^\ast$}}
  \put(2,4){\makebox(0,0){$\otimes$}}
  \put(3,4){\makebox(0,0){$H_{0234}$}}
  \put(4,4){\makebox(0,0){$\otimes$}}
  \put(5,4){\makebox(0,0){$H_{0134}^\ast$}}
  \put(6,4){\makebox(0,0){$\otimes$}}
  \put(7,4){\makebox(0,0){$H_{0124}$}}
  \put(8,4){\makebox(0,0){$\otimes$}}
  \put(9,4){\makebox(0,0){$H_{0123}^\ast$}}

  \put(1,1){\makebox(0,0){$H_{2134}$}}
  \put(2,1){\makebox(0,0){$\otimes$}}
  \put(3,1){\makebox(0,0){$H_{0134}^\ast$}}
  \put(4,1){\makebox(0,0){$\otimes$}}
  \put(5,1){\makebox(0,0){$H_{0234}$}}
  \put(6,1){\makebox(0,0){$\otimes$}}
  \put(7,1){\makebox(0,0){$H_{0214}^\ast$}}
  \put(8,1){\makebox(0,0){$\otimes$}}
  \put(9,1){\makebox(0,0){$H_{0213}$}}

  \put(1,3.5){\vector(0,-1){2}}
  \put(1.8,2.5){\makebox(0,0){${\tau_0^\ast}^{-1}$}}
  \put(3,3.5){\vector(1,-1){2}}
  \put(2.8,3){\makebox(0,0){$\id$}}
  \put(5,3.5){\vector(-1,-1){2}}
  \put(5.2,3){\makebox(0,0){$\id$}}
  \put(7,3.5){\vector(0,-1){2}}
  \put(7.4,2.5){\makebox(0,0){$\tau_1$}}
  \put(9,3.5){\vector(0,-1){2}}
  \put(9.8,2.5){\makebox(0,0){${\tau_1^\ast}^{-1}$}}

  \put(10,4){\vector(2,-1){2}}
  \put(11.6,4.2){\makebox(0,0){$Z_{01234}^{\prime (V),(+)}$}}
  \put(10,1){\vector(2,1){2}}
  \put(11.6,0.8){\makebox(0,0){$Z_{02134}^{\prime (\tilde V),(-)}$}}
  \put(12.5,2.5){\makebox(0,0){$\C$}}
\end{picture}
\end{center}
Employing Lemma~\ref{lemma_inverse}, this proves the claim.
\end{proof}

There exist similar lemmas for the other elementary transpositions
$(23)$ and $(34)$ as well as for the corresponding statements with
opposite relative orientations, \ie\ where $Z^{(V),(+)}$ and
$Z^{(V),(-)}$ are exchanged. Their proofs are entirely analogous.

\begin{theorem}
\label{thm_combinatorial}
The partition function~\eqref{eq_partribbon} does not depend on the
choice of the linear order of vertices.
\end{theorem}

\begin{proof}
Equip the set of vertices with a different linear order which is
induced from the given one by a permutation $\tau$ of the vertices.
The partition function using this new order can be expressed in terms
of the definitions of Section~\ref{sect_ribbondef} which use the original order, if
$\tau$ is applied both to the vertices and to the colouring,
\begin{equation}
\label{eq_partperm}
  Z_\tau = \sum_{V\colon\Lambda^{(2)}\to\CC_0}\biggl(
    \prod_{(v_0,v_1,v_2)\in\Lambda^{(2)}}\hat
    w(v_{\tau^{-1}(0)},v_{\tau^{-1}(1)},v_{\tau^{-1}(2)})({(\tau V)}_{012})\biggr)\, 
  Z_\tau^{(\tau V)}.
\end{equation}
Here
\begin{equation}
\label{eq_prelimperm}
  Z_\tau^{(\tau V)} = \tr_{\sym{H}} \biggl[P\circ
    \Bigl(\bigotimes_{s\in\Lambda^{(4)}} 
    Z^{(\tau V),(\epsilon_{\tau(s)})}_{\tau(s)}\Bigr)\biggr]
\end{equation}
replaces the partition function per colouring in the case of the new
vertex order, $\tau(s)$ denotes $(\tau^{-1}(0)\ldots\tau^{-1}(4))$ for
a given $4$-simplex $s=(01234)$, and $\tau V$ is the colouring induced
by $\tau$, \ie\ ${(\tau
V)}_{012}=V_{\tau^{-1}(0)\tau^{-1}(1)\tau^{-1}(2)}$ for all triangles
$(v_0,v_1,v_2)\in\Lambda^{(2)}$.

The permutation $\tau$ replaces triangles $(012)$ by
$(\tau^{-1}(0)\tau^{-1}(1)\tau^{-1}(2))$ and therefore just
permutes the factors of the product in~\eqref{eq_partperm}. This
product can be reorganized so that we obtain
\begin{equation}
\label{eq_partpermb}
  Z_\tau = \sum_{V\colon\Lambda^{(2)}\to\CC_0}\biggl(
    \prod_{(v_0,v_1,v_2)\in\Lambda^{(2)}}\hat
    w(v_0,v_1,v_2)(V_{012})\biggr)\, Z_\tau^{(\tau V)},
\end{equation}
where the vertex order of the triangles does not matter because of the
reality condition~\eqref{eq_reality}.

Any permutation $\tau$ which just permutes the $4$-simplices but does
not change the vertex order of these $4$-simplices, permutes the
tensor factors in~\eqref{eq_prelimperm} and therefore leaves the trace
invariant. It is thus sufficient to prove invariance under
permutations $\tau$ that change the vertex order for fixed
$4$-simplices. 

Consider a $4$-simplex $s=(01234)$ and and let $\tau$ be an elementary
transposition, $\tau\in\{(01),(12),(23),(34)\}$. The colouring $\tau
V$ associates with each triangle $(w_0,w_1,w_2)\in\Lambda^{(2)}$
either the object $V(w_3,w_4,w_5)$ assigned to some triangle
$(w_3,w_4,w_5)\in\Lambda^{(2)}$ or the dual of that object.

Since the set of colours $\CC_0$ contains for each given object $V$
exactly one object that is isomorphic to $V^\ast$, there exists a
unique colouring $\overline V\colon\Lambda^{(2)}\to\CC_0$ such that
${\overline V}_{012}\cong {(\tau V)}_{012}$ for all
triangles. Moreover since $\tau$ is a transposition, ${(\tau\overline
V)}_{012}\cong V_{012}$ so that $\tau$ induces an involution on the
set of colourings $\Lambda^{(2)}\to\CC_0$. We can now sum over
$\overline V$ rather than $V$ in~\eqref{eq_partpermb} and obtain
\begin{eqnarray}
  Z_\tau &=& \sum_{V\colon\Lambda^{(2)}\to\CC_0}\biggl(
    \prod_{(v_0,v_1,v_2)\in\Lambda^{(2)}}\hat
    w(v_0,v_1,v_2)(V_{012})\biggr)\, Z_\tau^{(V)},\\
  Z_\tau^{(V)} &=& \tr_{\sym{H}} \biggl[P\circ
    \Bigl(\bigotimes_{s\in\Lambda^{(4)}} 
    Z^{(V),(\epsilon_{\tau(s)})}_{\tau(s)}\Bigr)\biggr],
\end{eqnarray}
where we have used~\eqref{eq_reality} and where we have written $V$
instead of $\overline{V}$ for simplicity.

In the preceding lemmas we have constructed linear isomorphisms which
form the vertical maps in commutative diagrams of the following form:
\begin{center}
\setlength{\unitlength}{7mm}
\begin{picture}(13,6)
  \put(1,5){\makebox(0,0){$H_{0234}$}}
  \put(2,5){\makebox(0,0){$\otimes$}}
  \put(3,5){\makebox(0,0){$H_{0124}$}}

  \put(4,5){\vector(1,0){2}}

  \put(7,5){\makebox(0,0){$H_{1234}$}}
  \put(8,5){\makebox(0,0){$\otimes$}}
  \put(9,5){\makebox(0,0){$H_{0134}$}}
  \put(10,5){\makebox(0,0){$\otimes$}}
  \put(11,5){\makebox(0,0){$H_{0123}$}}

  \put(1,1){\makebox(0,0){$H^{(1)}$}}
  \put(2,1){\makebox(0,0){$\otimes$}}
  \put(3,1){\makebox(0,0){$H^{(2)}$}}

  \put(4,1){\vector(1,0){2}}

  \put(7,1){\makebox(0,0){$H^{(3)}$}}
  \put(8,1){\makebox(0,0){$\otimes$}}
  \put(9,1){\makebox(0,0){$H^{(4)}$}}
  \put(10,1){\makebox(0,0){$\otimes$}}
  \put(11,1){\makebox(0,0){$H^{(5)}$}}

  \put(1,4.4){\vector(0,-1){3}}
  \put(1.4,3){\makebox(0,0){$\Phi_1$}}
  \put(3,4.4){\vector(0,-1){3}}
  \put(3.4,3){\makebox(0,0){$\Phi_2$}}
  \put(7,4.4){\vector(0,-1){3}}
  \put(7.4,3){\makebox(0,0){$\Phi_3$}}
  \put(9,4.4){\vector(0,-1){3}}
  \put(9.4,3){\makebox(0,0){$\Phi_4$}}
  \put(11,4.4){\vector(0,-1){3}}
  \put(11.4,3){\makebox(0,0){$\Phi_5$}}
  \put(5,5.5){\makebox(0,0){$Z_{01234}^{(V),(+)}$}}
\end{picture}
\end{center}
Here the $H^{(j)}$ are suitable state spaces such that the bottom
horizontal map is related to the $4$-simplex map $Z^{(\tilde
V)(-)}_{\tau(s)}$ by the pairing~\eqref{eq_pairhom}. The colouring
$\tilde V$ is such that ${\tilde V}_{012}\cong V_{012}$ for all
triangles $(012)$.

Since each tetrahedron occurs twice in the boundary of some
$4$-simplices, once with positive and once with negative relative
orientation, each state space $H_{\tau(jk\ell m)}$ occurs twice among
the $H^{(j)}$, once as $H_{\tau(jk\ell m)}$ and once as the dual state
space $H^\ast_{\tau(jk\ell m)}$. In both cases, the corresponding map
$\Phi_j$ is the same, either one of the $\tau_i$ or the
identity. Therefore the tensor product of all $4$-simplex maps is
conjugated by a linear isomorphism $\Phi$ which can be obtained from a
tensor product of these $\Phi_j$,
\begin{equation}
\label{eq_undertraceb}
  P\circ\Bigl(\bigotimes_{s\in\Lambda^{(4)}}
    Z_s^{(V),(\epsilon_s)}\Bigr)
  = \Phi\,\Bigl[ P\circ\Bigl(\bigotimes_{s\in\Lambda^{(4)}} 
    Z_{\tau(s)}^{(V),(\epsilon_{\tau(s)})}\Bigr)\Bigr]\,\Phi^{-1}.
\end{equation}
Observe that here $\epsilon_{\tau(s)}=-\epsilon_s$ and that the
colouring $\tilde V$ can be replaced by $V$ as a consequence of
Theorem~\ref{thm_simple}. 

Since $Z_\tau^{(V)}$ is the trace over~\eqref{eq_undertraceb}, we find
$Z_\tau^{(V)}=Z^{(V)}$ and therefore $Z_\tau=Z$.
\end{proof}

\subsection{Gauge Symmetry}

In order to understand the gauge symmetry of LGT in the picture of the
Spin Foam Model, consider first the case in which $\CC$ is the
category of finite-dimensional representations of the gauge group
$G$. The group then acts on its representations via natural
equivalences ${(t_g^{(V)})}_V$, $g\in G$, \ie\ natural isomorphisms
$t_g^{(V)}\colon V\to V$ for all objects $V$.

In LGT on the $2$-complex $(V,E,F)$ dual to the simplicial complex
$\Lambda^{(\ast)}$, consider a gauge transformation involving only one
vertex $v\in V$. This means that the group elements $g_e$ attached to
the edges $e\in E$ are transformed as
\begin{mathletters}
\begin{alignat}{2}
  g_e&\mapsto g_e\cdot h_v^{-1},\qquad&\mbox{if}\quad v&=\del_-e,\\
  g_e&\mapsto h_v\cdot g_e,\qquad&\mbox{if}\quad v&=\del_+e,
\end{alignat}%
\end{mathletters}%
for $h_v\in G$ while all other variables $g_e$ remain unchanged. For
each polygon containing the vertex $v$ in its boundary, precisely two
edges are affected in such a way that the effect of the transformation
cancels for the polygon.

In the spin foam picture, only the $4$-simplex dual to the vertex
$v\in V$ is affected. Let $(v_0,v_1,v_2)$ denote the triangle to which
a polygon is dual. The gauge transformation then inserts natural
isomorphisms $t_g^{(V_{012})}$ and ${(t_g^{(V_{012})})}^{-1}$ into the
ribbon corresponding to that triangle, \ie\ to the ribbon labelled by
the object $V_{012}$. These isomorphisms cancel.

In the categorical picture, however, this symmetry can be understood in
other terms. Let now $\CC$ denote any admissible ribbon category and
choose a colouring $V\colon\Lambda^{(2)}\to\CC_0$. Consider a morphism
$\phi_{0123}\in H_{0123}$, \ie\ $\phi_{0123}\colon V_{123}\otimes
V_{013}\to V_{012}\otimes V_{023}$. Then for any natural equivalence
${(t^{(V)})}_V$, naturality means
\begin{equation}
  \phi_{0123} = (t^{(V_{013})}\otimes t^{(V_{012})})\circ
    \phi_{0123}\circ ({t^{(V_{123})}}^{-1}\otimes {t^{(V_{013})}}^{-1}).
\end{equation}
If this transformation for the natural equivalence ${(t^{(V)})}_V$ is
applied simultaneously to all morphisms $\phi_{jk\ell m}$ in
Figure~\ref{fig_vertex}(a) or~(b), the isomorphisms $t^{(V_{jk\ell})}$
cancel pairwise in each ribbon, and the quantum trace remains
unchanged.

In the categorical description of the Spin Foam Model, the gauge
symmetry is therefore automatically implemented. It is just the
naturality property of natural equivalences together with the fact
that all ribbon diagrams used in the definition of the partition
function are quantum traces.

\subsection{Wilson loop and spin networks}

Having generalized the partition function of LGT to ribbon categories,
it is desirable to understand the corresponding generalization for the
observables of LGT, namely for Wilson loops and spin networks
(Definition~\ref{def_spinnet}). 

In order to define the expectation value of a spin network, recall
that the quantum traces in Figure~\ref{fig_vertex} generalize the
four-dimensional version of Figure~\ref{fig_duality_partition}(a). One
should therefore extend Figure~\ref{fig_vertex} and include five
additional ribbons and one coupon for the spin network as
Figure~\ref{fig_duality_partition}(b) suggests. However, it seems to
be impossible to find a ribbon diagram which has the symmetries
required in Section~\ref{sect_welldef}.

A possible explanation is the following argument. In the Lie group
case, the spin network~\eqref{eq_spinnet} attaches representations to
the edges and morphisms to the vertices of the $2$-complex
$(V,E,F)$. Its generalization to the ribbon case should therefore be
given by a ribbon graph in four dimensions. In four dimensions,
however, there is no canonical way of associating to each ribbon graph
a morphism in the ribbon category $\CC$ because there is no
four-dimensional analogue of the Reshetikhin--Turaev functor. It is
conceivable that the notion of a spin network in four dimensions using
ribbon categories is not a good definition.

For the construction of observables that generalize
Definition~\ref{def_spinnet} to the case of ribbon categories, one has
therefore to choose a linear order of vertices on which the result
will then depend in a crucial way. Spin network observables are thus
defined for simplicial complexes, but not in general for combinatorial
complexes.

These restrictions are important if one wants to construct particular
physical models which are based on a Spin Foam Model using ribbon
categories. It remains to be studied under which conditions one can
define at least a certain class of observables and how the dependence
on the vertex order can be interpreted. The reader is also referred to
the diagrammatic approach to observables in three dimensions
in~\cite{Oe01}.

%
\section{Special cases and generalizations}
%
\label{sect_special}

The partition function~\eqref{eq_partribbon} of the Spin Foam Model
using ribbon categories covers a number of special cases which were
already known in other contexts. In this section, we comment on the
relations between these models.

\subsection{Lattice Gauge Theory}
\label{sect_special_lgt}

The category $\CC=\RCat G$ of finite-dimensional representations of a
compact Lie group $G$ forms a semi-simple admissible ribbon
category. The relation of the Spin Foam Model with LGT holds if the
set of colours $\CC_0$ is a set containing one representative of each
equivalence class of simple objects.

In this case one can use the generic Boltzmann
weight~\eqref{eq_boltzmann} for any action which is local, \ie\
evaluated once for each polygon, and which is a real and bounded
$L^2$-integrable class function of $G$. In particular, the standard
\emph{Wilson action} and the \emph{heat kernel} or \emph{generalized
Villain action} are of this form. For details about these actions and
about their character expansion, we refer the reader to standard
textbooks such as~\cite{MoMu94,Ro92}.

In general the set of representatives $\CC_0$ of the simple objects is
countably infinite. However, the partition
function~\eqref{eq_partribbon} is a convergent series because the
Boltzmann weight is an $L^2$-function, and its character expansion
therefore forms a square summable series due to the Peter--Weyl
Theorem. For more details, see~\cite{OePf01,CaSe95}.

In this case both pictures, LGT and the Spin Foam Model, are
well-defined and are dual to each other in the sense
of~\cite{OePf01}. A comparison of the Spin Foam Model dual to
LGT~\eqref{eq_dual_partition} and the
generalization~\eqref{eq_partribbon} shows the following
correspondences, \cf~Table~\ref{tab_dualcomplex}. The sum over
colourings of triangles/polygons and the weights $\hat w$ are
explicitly contained in the partition function. The sum over
colourings of tetrahedra/edges with morphisms is explicit
in~\eqref{eq_dual_partition} and it is the result of the trace over
the tensor product $\sym{H}$ in~\eqref{eq_prelimpart}. The weights
$C(v)$ per $4$-simplex/vertex are given by the formula~\eqref{eq_cv}
and agree with the quantum traces of Figure~\ref{fig_vertex} which
appear as a result of the trace over the $4$-simplex maps
in~\eqref{eq_prelimpart}.

For standard actions of LGT such as Wilson's action or the heat kernel
action, the character expansion coefficients behave qualitatively like
$\exp(\frac{1}{\beta}s^\ast(V_\rho))$ if the Boltzmann weight is of
the form $\exp(\beta s(g))$. Here $\beta$ is the \emph{inverse
temperature}, $s(g)$ denotes the action and $s^\ast(V_\rho)$ the
\emph{dual action}, a function assigning a real number to each
finite-dimensional irreducible representation of $G$. The
transformation between LGT and the Spin Foam Model thus realizes a low
temperature --- high temperature duality or a strong-weak duality in
the bare coupling $g_0$ if $\beta=1/g_0^2$.

For the heat kernel action, the dual action $s^\ast(V_\rho)\sim
C^{(2)}_\rho$ is essentially given by the second order Casimir
operator $C^{(2)}_\rho$ of the representation. One can thus sort the
configurations of the Spin Foam Model by the sum of the Casimir
eigenvalues over all triangles, and recovers the full strong coupling
expansion of non-Abelian LGT.

Observe finally that LGT was formulated here on the $2$-complex dual
to a generic triangulation. In order to obtain the usual continuum
limit, the Boltzmann weight $\hat w(v_0,v_1,v_2)$ should now depend on
the geometry of the triangle $(v_0,v_1,v_2)$ in a suitable way.

\subsection{Gauge theory with finite groups}

If $\CC=\RCat G$ is the category of representations of a finite group
$G$, all comments of Section~\ref{sect_special_lgt} still apply. In
this case, there are only finitely many simple objects up to
isomorphism and the partition functions in both pictures, in LGT and
in the Spin Foam Model, are well-defined. 

It is now also possible to study the `topological' Boltzmann
weight
\begin{equation}
  w(g)=\delta (g):=\left\{
    \begin{array}{ll}
      |G|,&\mbox{if}\quad g=1,\\
      0   &\mbox{else}
    \end{array}\right.
  \qquad\mbox{\ie}\qquad \hat w_\rho=\dim V_\rho.
\end{equation}
With suitable prefactors, the partition function is then independent of
the triangulation and thus forms a topological invariant which is
well known and depends only on the gauge group and on the first
fundamental group of the manifold. See, for example, the comments in
Section~2.2 of~\cite{Bo97}.

\subsection{The Crane--Yetter state sum}

Let $\CC$ be a finitely semi-simple and admissible ribbon category
satisfying the conditions of Corollary~\ref{corr_peterweyl} and
$\CC_0$ be a set containing one representative for each equivalence
class of simple objects. This case is beyond the standard formulation
of LGT, and only the Spin Foam Model~\eqref{eq_partribbon} makes
sense. The partition function is a finite sum. It is again possible to
choose `topological' Boltzmann weights which here means the quantum
dimension of the simple objects,
\begin{equation}
  \hat w(V)=\qdim V.
\end{equation}
With suitable prefactors the partition function~\eqref{eq_partribbon}
agrees with the Crane--Yetter invariant~\cite{CrKa97}. For a comparison
of Figure~\ref{fig_vertex}(a) with the main diagram in~\cite{CrKa97},
observe that the state spaces $H_{0123}$ used in the present paper can
be further decomposed employing semi-simplicity~\eqref{eq_petercateg},
for example,
\begin{equation}
\label{eq_comparecrye}
  H_{0123}=\Hom(V_{123}\otimes V_{013},V_{012}\otimes V_{023})\cong
  \bigoplus_{J\in\CC_0}\Hom(V_{123}\otimes V_{013},J)\otimes\Hom(J,V_{012}\otimes V_{023}).
\end{equation}
If this decomposition is applied to the state spaces associated with
all tetrahedra, one has to colour in addition the tetrahedra with
simple objects (the $J$ in~\eqref{eq_comparecrye}) and 
the tetrahedra $(0123)$ with two types of morphisms,
$\Hom(V_{123}\otimes V_{013},J)$ and $\Hom(J,V_{012}\otimes
V_{023})$. These colourings are used in the standard formulation of
the Crane--Yetter state sum in~\cite{CrKa97}. Note that the
additional weight $1/\qdim J$ per tetrahedron in~\cite{CrKa97} is a
consequence of the choice of bases of $\Hom(V_{123}\otimes V_{013},J)$
and $\Hom(J,V_{012}\otimes V_{023})$.

\subsection{The generic case}

The construction presented generalizes LGT and the Crane--Yetter
state sum, but also contains the generic case. Here $\CC$ is any
admissible ribbon category, in particularly not required to be
semi-simple, and the weights $\hat w(V)$ for given simple objects $V$
can be quite freely chosen. If the set of colours $\CC_0$ is finite,
the partition function is a finite sum and thus well-defined for any
choice of weights. If $\CC_0$ is a countable set, similar convergence
issues arise as for Lie groups~\cite{OePf01}. Note that here it is
also necessary to examine the quantum traces of
Figure~\ref{fig_vertex} in order to prove convergence of the partition
function.

\subsection{Generalizations and the Barrett--Crane model}

If $\CC=\RCat G$ for a compact Lie group, for example, $G=\SU(2)$, and
if the Boltzmann weight is chosen to be `topological',
\begin{equation}
  w(g)=\delta (g),\qquad \hat w(V_\rho)=\dim V_\rho,
\end{equation}
the partition function is just a (divergent) formal expression. This
is the case for the Ooguri model~\cite{Oo92} which can be formulated
in the LGT or in the spin foam picture.

The simplest version of a Spin Foam Model of Barrett--Crane
type~\cite{BaCr98} is obtained from the Ooguri model for
$\SO(4)$ in the spin foam picture by restricting the
representations in all sums to the \emph{simple} representations of
$\SO(4)$. Simple here means that the representation is of the form
$V\otimes V$ as a representation of $\SU(2)\times\SU(2)$ for some
irreducible representation $V$ of $\SU(2)$.

In order to implement this restriction one can choose the set of
colours $\CC_0$ to contain one representative per isomorphism class of
simple representations of $\SO(4)$. However, in addition one has to
restrict the sum over $J$ in~\eqref{eq_comparecrye} to simple
representations. As a consequence the state spaces $H_{0123}$ are
certain subspaces of $\Hom(V_{123}\otimes V_{013},V_{012}\otimes
V_{023})$.

The results of the present paper can be generalized to state spaces
that are subspaces of $\Hom(V_{123}\otimes V_{013},V_{012}\otimes
V_{023})$ as long as the pairing~\eqref{eq_pairhom} and the maps
$\tau_j$ of Definition~\ref{def_snaction} can be consistently
restricted to these subspaces. The correspondence with LGT with a
partition function~\eqref{eq_partfunc} is, however, lost as a
consequence of this generalization.

%
\section{Conclusion and Outlook}
%
\label{sect_outlook}

The Spin Foam Model for ribbon categories defined in the present paper
generalizes the Spin Foam Model dual to Lattice Gauge Theory (LGT) and
can be used as a definition of LGT for gauge groups which are quantum
groups rather than Lie groups. Furthermore the definition presented
here encompasses state sum models that are of interest both in
topology and in quantum gravity. The definition presented provides a
bridge between the standard (Lie group) formulation of LGT and the
Crane--Yetter invariant which uses ribbon categories. It can also be
used to construct other Spin Foam Models that do not correspond to
Topological Quantum Field Theories and provides proofs that they are
well-defined. This work might finally help to make the relation of LGT
and the Spin Foam Models used in other areas more transparent and the
common concepts and open questions more accessible.

If one seeks to construct even more general Spin Foam Models than
defined here, it is worth pointing out that consistency of the
definition restricts the quantum traces of Figure~\ref{fig_vertex} very
tightly. The introduction of further weights, however, seems to be
much easier to achieve.

The definition of LGT with ribbon categories presented here is
restricted to $4$ dimensions. Technically, this is due to the fact
that the key diagrams in Figure~\ref{fig_vertex} are hand made for
this construction. Due to the generality of ribbon categories it
involves choices of over- or under-crossings, and only with a good
choice is the partition function well-defined. While the corresponding
approaches in $d=3$ \cite{Bo93,Oe01} are canonical in the sense that
their construction is well-defined due to general principles, the
$d=4$ construction presented here involves choices and one has to
verify {\sl a posteriori\/} that it is consistent. It is not obvious
whether the result of~\cite{CaCa01} in arbitrary dimension in the Lie
group case can be generalized to ribbon categories. It is, in any
case, a striking observation that there exist constructions in
$d=3$~\cite{Oe01} and in $d=4$ (presented here) which both generalize
the Spin Foam Model dual to LGT. Notice, however, that the $d=3$ case
can be handled with spherical categories~\cite{BaWe96} which are more
general than ribbon categories. For the construction in $d=4$, we make
explicit use of ribbon categories because the basic diagrams in
Figure~\ref{fig_vertex} always contain a crossing. If ribbon
categories are used in the construction in $d=3$, all spin network
observables can be defined~\cite{Oe01} whereas in $d=4$ there are no
canonical expressions for the spin network observables anymore.  It
therefore seems that one has to use more and more restrictive
structures if one wishes to increase the dimension.

\acknowledgements

The author is grateful to the German Academic Exchange Service DAAD
for his HSP~III scholarship. I would like to thank R.~Oeckl for
stimulating discussions, in particular for many suggestions concerning
the categorical framework, and for providing me with preliminary
versions of~\cite{Oe01}. I~am furthermore grateful to
A.~J.~Macfarlane, S.~Majid, D.~Oriti and R.~M.~Williams for
discussions, for comments on the manuscript and on relevant
literature.

%
%

\begin{figure}[ht]
\begin{center}
\input{fig/vertex_apply_01.pstex_t}
\mycaption{fig_vertex_apply_01}{%
  This diagram is isotopic to the quantum trace
  $Z^{\prime(V),(+)}_{01234}$ in Figure~\ref{fig_vertex}(a), \cf\ 
  Lemma~\ref{lemma_perm01} and Lemma~\ref{lemma_commutative01}. The
  morphisms in the dashed boxes are by definition of $\tau_0$
  (Definition~\ref{def_snaction}) just morphisms
  $\overline\phi_{1024}\in H_{1024}^\ast$, $\phi_{1034}\in 
  H_{1034}$ and $\phi_{1023}\in H_{1023}$. With these replacements
  this quantum trace is similar to Figure~\ref{fig_vertex}(b) defining
  $Z^{(V),(-)}_{10234}$ for opposite relative orientation with a
  non-standard order $(10234)$ of vertices. Note that the permutation
  $(01)\in\S_5$ also replaces $V_{012}$, $V_{013}$ and $V_{014}$ by
  their duals according to Definition~\ref{def_colouring}.}
\end{center}
\end{figure}

\begin{figure}[t]
\begin{center}
\input{fig/vertex_apply_12.pstex_t}
\mycaption{fig_vertex_apply_12}{%
  This diagram is isotopic to the quantum trace defining
  $Z^{(V),(+)}_{01234}$ in Figure~\ref{fig_vertex}(a), \cf\
  Lemma~\ref{lemma_perm12} and Lemma~\ref{lemma_commutative12}.}
\end{center}
\end{figure}

\end{document}